\documentclass{WP_template_Hyp}[12pt]

\title{JaxNet: Scalable Blockchain Network}
\author[1]{Iurii Shyshatsky \thanks{iurii\@jax.net}}
\author[1]{Vinod Manoharan \thanks{vinod\@jax.network}}
\author[1]{Taras Emelyanenko \thanks{taras\@jax.network}}
\author[1]{Lucas Leger \thanks{lucasyleger\@gmail.com}}
\affil[1]{JaxNet, Kyiv, Ukraine %\\\href{ourmail}{\texttt{ourmail}}
}
\date{\today}

\begin{document}

\pagestyle{fancy}
\pagenumbering{arabic}

\maketitle

\begin{abstract}
Today’s world is organized based on merit and value. A single global currency that’s decentralized is needed for a global economy. Bitcoin is a partial solution to this need, however it suffers from scalability problems which prevent it from being mass-adopted. Also, the deflationary nature of bitcoin motivates people to hoard and speculate on them instead of using them for day to day transactions. We propose a scalable, decentralized cryptocurrency that is based on Proof of Work. The solution involves having parallel chains in a closed network using a mechanism which rewards miners proportional to their effort in maintaining the network. The proposed design introduces a novel approach for solving the scalability problem in the blockchain network based on merged mining.
\end{abstract}

%\begin{keywords}
%\end{keywords}

%\tableofcontents

\section{Introduction}
\subsection{Need for a global decentralized currency.}
Governments, through their monetary policies, influence the value of a currency. A person could work for her entire life, save a part of her wealth in their currency and lose everything overnight when the currency collapses due to bad monetary policy. 

If this is less probable to occur in well manage countries with sound institutions, bad money is still a fact. People from some developing nations, although very productive, are still unable to create value for themselves as much as people from developed nations due to unstable local governments. Effective competition remains at the local level, and restricted within currency markets. 

In other words, we have local currencies in a globalized world. A single global decentralized currency would create a fairer and more competitive environment, removing all the inequalities created by their respective governments. This would in turn increase the overall productivity of our civilization and speed up advancement in all spheres.
\subsection{Paper organization}
This paper is organized as follows. In the \cref{ScTrilemma} a brief introduction to the Blockchain Scalability Problem is given. In the \cref{secArc} the design of our solution is discussed. Brief description of the main idea is given in the \cref{MainIdofSol}. \cref{secEco} mining reward scheme is described and how it affects the network economy. Network security is discussed in the \cref{SecMod}. Finally, \cref{secCon} contains discussion of the results and open problems.
%\section{Background}
\subsection{Scalability of the distributed network} \label{ScTrilemma}

After the launch of the Bitcoin network \cite{Satoshi}, we witness with its exponential capitalization growth a Cambrian explosion of new altcoins. Naturally, a new question arose: could BTC (or any of its numerous rivals), become an universal global currency with cheap maintenance? Could it be made fast and acceptable even for small transactions in any gas-station, fast-food, or apparel shop, as well as secure and reliable for big transactions between large corporations? What about big, complicated, long term contracts like those for cargo ships, opening a credit line, or setting supply chain contracts? 

Among the leading problems are legislation, the position of the financial market's regulators, and the attitude of individual suppliers and consumers. However, there are more technical obstacles on the way to mass adoption. All together, these obstacles are commonly called \textit{scalability problem in distributed system}. Another name for this problem is \textit{Scalability Trilemma}. This term was introduced by the co-founder of the Ethereum project Vitaliy Buterin. \cite{samani2018models,MediumTrilemma,EthSharding}

\begin{figure}[ht!]
	\begin{center}
		\resizebox{\textwidth}{!}{
			\begin{tikzpicture}[
			%blend group=screen,
			%fill opacity=0.8,
			%path fading=fade down,
			]
			%[mypicture,auto]
%\draw[draw=white] (0,0) circle(50pt);
%\shade[shading = color wheel] [even odd rule]
%\pgfsetopacity{0.5}
%\filldraw[top color=red,bottom color=white] (0,0) circle (3);
%\filldraw[left color=green,right color=white] (0,0) circle (3);
\clip (-4,-1.5) rectangle (4,2);
\begin{scope}[transparency group]
\clip[rotate=30] (-2,-1.8) rectangle (1,1.8);
\clip[rotate=-30] (-1,-1.8) rectangle (2,1.8);
\clip (-2,-1) rectangle (2,2);
\foreach \x in {0,0.0111,...,1} {
	\definecolor{currentcolor}{hsb}{\x, 0.8, 0.9}
	\draw[draw=none, fill=currentcolor]
		(-360*\x+28:0.2) -- (-360*\x+88:2)
		-- (-360*\x+32:2) -- (-360*\x+92:0.2) -- cycle;
}
\end{scope}
%\begin{scope}[transparency group]
%\begin{scope}[blend mode=screen]
%\fill[left color=white,right color=red,rotate=60,
%path fading=west,rotate=30,fading angle=30
%] (-2,-1.8) rectangle (1,1.8);
%\fill[left color=green,right color=white,rotate=90,
%green,path fading=east,rotate=-30,fading angle=-30
%] (-1,-1.8) rectangle (2,1.8);
%\fill[blue,path fading=north] (-2,-1) rectangle (2,2);
%\end{scope}
%\end{scope}
%\fill[green,path fading=west,rotate=60,fading angle=180] (0,0) circle (2);
%\fill[blue,path fading=west,rotate=120,fading angle=120] (0,0) circle (2);
%\node[TrilemmaTr] (Tr) at (0,0) {};
\node[TrilemmaCirc,
label = {[label distance=5pt, yshift=17pt, rotate=-60] 60: Scalability},
label = {[label distance=9pt] below: Decentralization},
label = {[label distance=5pt, yshift=17pt, rotate=60] 120: Security}
] (Circ) at (0,0) {?};
			\end{tikzpicture}}
		\vspace*{3pt}
		%\captionsetup{labelformat=empty}
	    \caption{\textbf{Blockchain Scalability Trilemma}} \label{Trilemma}	
    \end{center}
\end{figure}
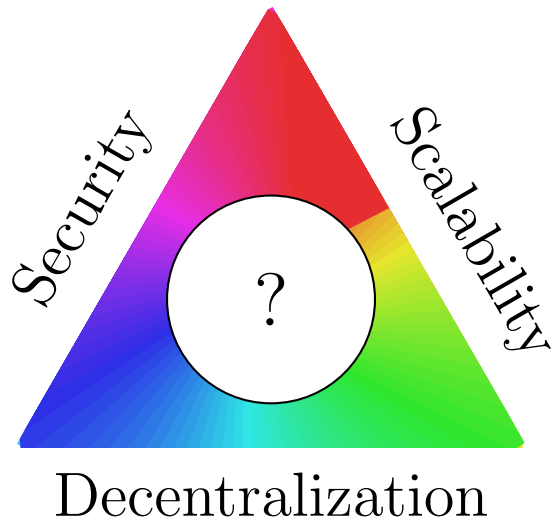

When the size of the distributed system grows it requires more maintenance. If a given blockchain network becomes more global, then it should, on the one hand, have billions of wallets and execute billions of transactions every day. So, it should be rather powerful to handle these transactions throughout and store part of the transaction history required for thorough verification. On the other hand, this system should be friendly enough for small nodes to help maintain the network with their limited resources. 

Such a system should be convenient for users in terms of short confirmation time, easy and reliable transactions, and with low fees. Overall, distributed networks should be decentralized, and thus be an alternative to classical solutions of the problem through decentralization. Even though blockchain technology widely uses some sophisticated techniques to record, maintain, and validate data like encryption, hashes, and Merkle trees, there are still significant limitations. 

First, any node wanting to operate as a full node has to track everything that is going on in the network. He has to know whether somebody signed a transaction in order to buy a cup of coffee and keep the history of such transactions for a significant amount of time. Many nodes operate only as light nodes and rely heavily on full nodes. 

The more global the network becomes, the worse it gets. Only supercomputers and fiber-optic communications can do the job efficiently. Personal devices like smartphones and laptops are too far away from such performance in the foreseeable future. 

This means more centralization, and only the biggest nodes will survive, defeating the case of centralization in the first place. Scalability has its cost, but we are about to choose how to pay for it. The classical solution of the scalability problem is centralization. Many projects set a goal to find a solution that avoids centralization.\cite{zhou2020solutions} We claim that we have found a design which is better than existing solutions, and we are going to describe it in this paper.

\section{Architecture}\label{secArc}
\subsection{Solution Overview}
\subsubsection{Concepts and notations}
In this section, we are going to discuss key concepts relevant to Jaxnet that are used throughout the paper. We assume that the reader is familiar with the concept of a distributed network and the basic blockchain design of Bitcoin introduced by Satoshi Nakamoto. Since Jaxnet design has multiple differences from Bitcoin we have to review common concepts in the new setting and introduce some new ideas. In short, unlike many other blockchain systems, Jaxnet maintains multiple chains of blocks, use specific mining and reward schemes. To avoid confusion we need a clear description of this setting. Moreover, blockchain is rather new technology and the set of common notations and concepts is not well established. Frequently word meaning depends on the context.

%\begin{definition}
\textit{Node} is a basic component of the distributed network. Nodes constitute the distributed network.
%\end{definition}
%\begin{remark}
Nodes are often divided into two types: \textit{full nodes} and \textit{light nodes}. Full-nodes perform necessary communication and store necessary data to verify data on their own. Light nodes reduce their storage and network communication expenses. Thus they rely in some way on full nodes and accept respective security risks. One of the goals of scalable distributed network protocol is to reduce such expenses and allow more nodes to operate as full-nodes.  
%\end{remark}

%\begin{definition}
\textit{Blockchain} is the replicated database in the form of blocks of records linked into chains.\cite{zhao2020algebraic}
%\end{definition}
Sometimes the whole network which maintains chains of blocks is called the blockchain.
%\begin{definition}
%JaxNet is a distributed network described in this paper. 
%\end{definition}
%\begin{definition}\
%(rewrite)
\textit{Mining} is the process of selecting a leader who adds new data to the large distributed network known as the blockchain network through the consensus.
%\end{definition}
%\begin{remark}
Although the Bitcoin blockchain uses the consensus protocol based on solving a hard puzzle known as "Proof of Work" some other approaches were developed.\cite{xiao2020survey}
%\end{remark}
%\begin{definition}
\textit{Proof of Work} or \textit{PoW} is a type of consensus protocol in which miners rich consensus by participating in the contest of solving hard puzzles. 
%\end{definition}
Like Bitcoin blockchain, Jaxnet uses the Proof of Work approach. However, in contrast to Bitcoin Jaxnet maintains multiple blockchains and miners participate in multiple contests simultaneously. Miners have the freedom of choice in what subset of blockchains to mine and, respectively, what subset of PoW contests to participate in. Therefore in the setting of JaxNet, we have the following definition.
%\begin{definition}
\textit{Mining in the JaxNet} is the process of selecting a leader who commits new data to the blockchains by reaching the PoW consensus described in this paper.
%\end{definition}

An important concept in JaxNet is "shard". Although this word is widely used we will set its meaning in the context of JaxNet, and use the concept through this paper.
%\begin{definition}
A \textit{ shard} in JaxNet is the collection of nodes that participate in each shard together with the shard-blockchain maintained by this collection. Shards constitute JaxNet. The number of shards is regulated by the JaxNet protocol. Any node can participate in as many shards as he wants, as long as he poses enough storage and bandwidth resources.
%\end{definition}
The maximal number of shards in JaxNet is not fixed.  Let us denote $N$ to be the number of shards in JaxNet at some moment. Besides shard-chains, the JaxNet protocol maintains one special chain. Let us call it the Beacon Chain or BC in short. Therefore, there are $N+1$ chains in JaxNet. The number of shards $N$ could be increased based on the consensus protocol conducted on the Beacon chain.

Blocks in the blockchain networks typically consist of two parts: block header and block body. This design improves the throughput of information across the network. Typically the \textit{block header} contains the most important data. It includes the reference to the body of the block and the hash of the previous header block in the chain. 

Block headers are often used to set the order of blocks in the chain and validate the block body data. Therefore, the chain of block headers without bodies is often called a blockchain. \textit{Block bodies} contain information about the transactions. Block content is discussed in the \cref{MMScheme}

In JaxNet there are two types of blocks: blocks in the Beacon Chain and blocks in shard chains. In JaxNet both types of blocks contain the header part and the body part.

During the mining process a miner proposes to extend chains with his versions of blocks. The blocks in his proposal are aggregated in the specific structure used for merged mining discussed in the \cref{MMproof}. The motivation of the miner is the reward. In Jaxnet this reward consists of two parts and described in the \cref{SetReward}.

Some recently established blockchain networks achieve good performance at the cost of decentralization. In these systems transaction verification, block generation, and acquiring network rewards, are restricted to the group of nodes called \textit{validators}. Therefore, not all nodes in the network are the same and nodes often need a permission to become a validator. 

In contrast to the aforementioned systems, Jaxnet is a permissionless \textit{peer-to-peer} network. Nodes are equal and have identical roles similarly to Bitcoin design. No permission is needed to join the network.%Network maintenance and rewards are split between participants.
\subsubsection{Sharding of the distributed network}\label{sharding}
\begin{figure}[ht!]
	\begin{center}
		\resizebox{\textwidth}{!}{
			\begin{tikzpicture}[mypicture,auto]
	\foreach \x/\y/\z in 
	    {
	    A1/10/10,
	    A2/10/14,
	    A3/13/15,
	    A4/14/11
	    } 
	{
        \node at (\y,\z) (\x) [ShardingNodeGrey] {\x};
        \def\u{U\x}
        \node at (\y+30,\z) (\u) [ShardingNodeR] {\x};
    }
    \foreach \x/\y/\z in 
	    {
	    B1/0/10,
	    B2/0/14,
	    B3/3/15,
	    B4/4/11
	    } 
	{
        \node at (\y,\z) (\x) [ShardingNodeGrey] {\x};
        \def\u{U\x}
        \node at (\y+30,\z) (\u) [ShardingNodeG] {\x};
    }
    \foreach \x/\y/\z in 
	    {
	    C1/0/0,
	    C2/0/4,
	    C3/3/5,
	    C4/4/1
	    } 
	{
        \node at (\y,\z) (\x) [ShardingNodeGrey] {\x};
        \def\u{U\x}
        \node at (\y+30,\z) (\u) [ShardingNodeB] {\x};
    }
    
    \foreach \x/\y/\z in 
	    {
	    D1/10/0,
	    D2/10/4,
	    D3/13/5,
	    D4/14/1
	    } 
	{
        \node at (\y,\z) (\x) [ShardingNodeGrey] {\x};
        \def\u{U\x}
        \node at (\y+30,\z) (\u) [ShardingNodeO] {\x};
    }
    \node[MyShard,fit=(A1) (A2) (A3) (A4)
                      (B1) (B2) (B3) (B4)
                      (C1) (C2) (C3) (C4)
                      (D1) (D2) (D3) (D4),
        label={[black]90:{\scalebox{1.4}{\huge Unsharded Network}}}]  {};
	\node[MyShard,draw=red,fit=(UA1) (UA2) (UA3) (UA4),
        label={[black]90:{\scalebox{1.4}{\huge Shard A}}}] {};
	\node[MyShard,draw=green,fit=(UB1) (UB2) (UB3) (UB4),
        label={[black]90:{\scalebox{1.4}{\Huge Shard B}}}] {};
	\node[MyShard,draw=blue,fit=(UC1) (UC2) (UC3) (UC4),
        label={[black]90:{\scalebox{1.4}{\huge Shard C}}}] {};
	\node[MyShard,draw=orange,fit=(UD1) (UD2) (UD3) (UD4),
        label={[black]90:{\scalebox{1.4}{\huge Shard D}}}] {};
	
    \foreach \P/\Q/\C in 
	    {
	    A3/A1/black,
	    UA3/UA1/red,
	    C1/C3/black,
	    UC1/UC3/blue,
	    C1/C4/black,
	    UC1/UC4/blue,
	    B1/B2/black,
	    UB1/UB2/green,
	    D3/A1/black,
	    A1/B3/black,
	    B4/D1/black} 
	{
        \draw [myarrow,draw=\C!70] (\P) -- (\Q);
    }
    \shadedraw[->,
    shading=axis,
    shading angle=90,
    draw=gray!35, 
    %double=gray!20, 
    %double distance=20mm, 
    line width=20mm,
    {}-{Latex[length=25mm,width=60mm]}] (17,7.5) -- (27,7.5);
			\end{tikzpicture}}
		\vspace*{3pt}
		%\captionsetup{labelformat=empty}
	    \caption{\textbf{Network Sharding}} \label{Sharding}	
    \end{center}
\end{figure}
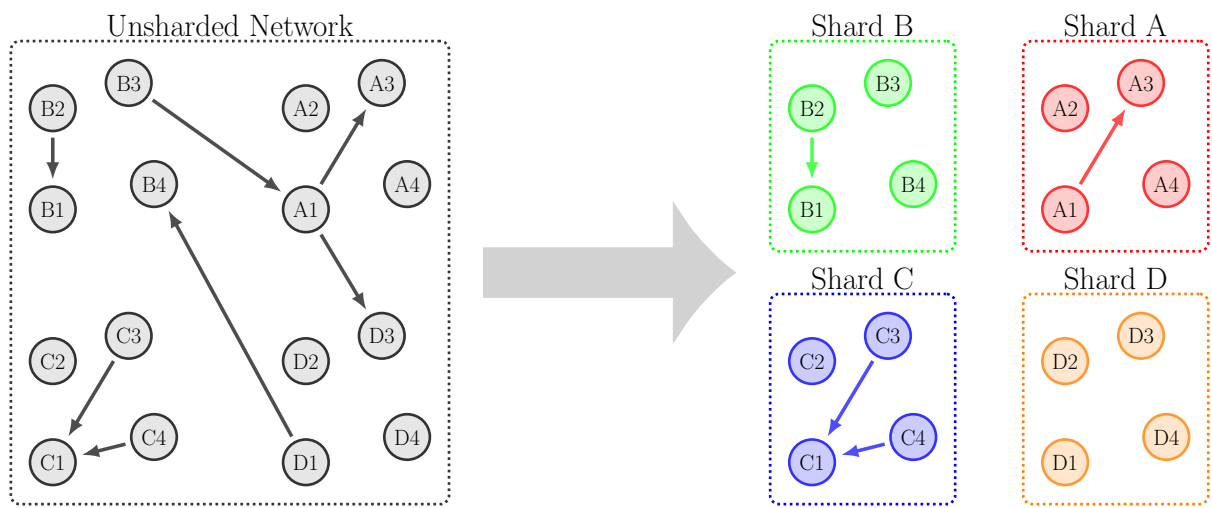

One of the possible approaches to the blockchain scalability problem is sharding (figure \ref{Sharding}).

The use of the term \textit{sharding} could be traced to the problem of database scaling.\cite{DBshard} Sharding arose as a natural approach for solving it.

Sharding of the blockchain network could bring benefits and disadvantages.\cite{ShardingBlockch,EthSharding} One argument in favour of sharding is that it could reduce the amount of information that a single node should process and store. This is a crucial requirement for a scalable blockchain network. 

However, this approach causes two problems. First, there is a need for a convenient and reliable tool for transferring funds between accounts in different shards. Second, the solution should preserve the same level of security. In particular, it should have the same or equivalent level of the resistance to 51\%-attacks for each shard in the network as could be achieved by a single network with the hash rate equal to an aggregate hash rate of every shard. Our solution addresses these problems.
\subsubsection{Merged mining}

Merged mining is a technology of particular importance in our design. It is a process of mining two or more cryptocurrencies at the same time, without sacrificing overall mining performance.\cite{MMbyBinance,zamyatin2016merged} In simple terms, if one compares generating hashes during mining with rolling a dice and the blockchain network with some entity which rewards you whenever the correct number is rolled, then with merge mining you get rewarded simultaneously from multiple entities while rolling the same dice. Merge mining was first described by Satoshi Nakamoto on the aforementioned forum in 2008. His idea was first implemented in Namecoin and then was used in the list of other projects. \cite{MMbyCC,MMbyBTC,CSbinance,zamyatin2016merged}

There are multiple approaches to organize merged mining. One has to follow one principle though; merged mined coins always share the same hashing algorithm. Miners collect profit from mining both, or more, chains. However, this profit is not necessarily the same on both chains. In the pair Bitcoin and Namecoin most of the profit is collected from the Bitcoin chain. In the pair Dogecoin \& Litecoin profits are closer to each other. In the second case, hash rates show significant correlation with each other since miners often try to mine both of them to maximize their profits.\cite{CSbinance,zamyatin2016merged} In some cases, merged mining becomes unprofitable since bandwidth and data storage expenses on one of chains exceed revenue.

Merged mining was already proposed as a possible approach to solving the Scalability Trilemma. However, observers said the benefits of merged mining could be achieved in more efficient way by setting just one chain with larger block size. Sometimes merged mining is considered to be a stealth block size increase.
\subsubsection{Nonstandard block reward allocation schemes}
%JaxNet design aims to make a step forward in this field.
%bring innovation to this field.

As December of 2019, the majority of cryptocurrency projects follow block reward and coin issuance rules similar to the Bitcoin design. Some projects like Ethereum don't have predefined rules. Although there were proposals of so-called \textit{algorithmic stablecoins} \cite{StableCoin2019}. 

Their history could be traced back to the \textit{Bitcointalk} forum that was used by Bitcoin community. In particular, some participants proposed "stablecoins pegged to the electricity". Proposals such as Encoin, GEM, Inertiacoin, Decrits were discussed there in 2011-2014. 

However, these proposals haven't withstood criticism, and to the best of our knowledge, were never finalized. Nevertheless, similar proposals sometimes appear in non-academic publications.\cite{CottenPost2018}%The author of this paper has found that similar ideas were discussed by Tim Cotten in the blog post \cite{CottenPost2018}.

Unlike the majority of blockchain protocols, JaxNet doesn't follow the "fixed block reward" rule. Instead of that the block reward is proportional to the block difficulty. 

However, in contrast to aforementioned blockchain proposals JaxNet doesn't have a goal of designing a stablecoin. It's exchange rate will float according to market demand and supply. 
 
Since the block reward rule used in JaxNet is not common there is no deep research devoted to it. Nevertheless, in the recent paper by Chen et. al. \cite{chen2019axiomatic} there is a study of reward allocation schemes and requirements which should be satisfied in order to make blockchain network to properly work. 

The authors of that paper determine a narrow class of such schemes and lists few examples. Remarkably, the reward scheme in JaxNet falls into the category \textit{generalized proportional allocation rules} and listed as "Example 3". Simultaneously the similar result was achieved by Leshno and Strack.\cite{leshno2019bitcoin} It's possible that there were other proposals, however, we are not familiar with them.
%Also, it is worth mentioning that due to various reasons experiments within this field of research often get little credit from the community.   
\subsubsection{Main idea of the solution} \label{MainIdofSol}
%In order to raise attention to the present paper the author is going to discuss the main idea of the solution here at the very beginning.

Sharding and merge-mining have already been proposed by researchers as possible approaches for the blockchain scalability problem, along with many ideas. However, as it was observed, the simultaneous usage of them in one solution causes a centralization problem. Since during merge mining, a miner can participate in merge mining of multiple chains simultaneously he can collect rewards from all of them, Therefore, if he calculated $h$ hashes within some time interval $t$ then his mathematical expectation of his reward $R_{total}$ is a sum over expected rewards $R_i$, $i=\overline{1,n}$ in every chain he has been mining:
\begin{equation}\label{exprew1}
    \Extn(R_{total}(h)) = \Extn(R_1(h))+\Extn(R_2(h))+...+\Extn(R_n(h))
\end{equation}
The equation \ref{exprew1} implies that the more shards a miner mines the more rewards he is about to get, and miners who mine more shards get a huge advantage over those who have limited resources. Therefore all miners are very interested in mining as many shards as possible. A more partitioned network become the less profitable one for weak players. 

Simply put, they do not mine enough coins to pay for their electricity bills. As the scaling continues the more centralized a network becomes. All remaining players in it mine nearly every shard. So, in this case, we are about to lose the decentralization part of the Scalability Trilemma. Therefore Vitaliy Buterin in his discussion of the problem compared merge mining to stealth increasing of the block size.\cite{EthSharding}  

This argument looks concrete. However, the authors of this paper think there is a flaw in its rationale. For example, let us split the reward into two parts: issued coins and collected transaction fees. Let's talk about issued coins and leave transaction fees apart for a while.
\begin{itemize}
    \item 
First, assume that the reward for mining the block in a shard depends on the number of shards which its builder was mining.
    \item
Second, assume that everybody in the shard can learn this number from the information printed in the block without any knowledge of what is going on in other shards.
    \item
Third, assume that reward functions in every shard are designed in such a way that the expected reward in each shard is proportional to one's effort or average hash rate. 
    \item
Finally, assume that coins in each shard have nearly the same value.
\end{itemize}
Then one can compute:
\begin{equation}
    \Extn(R_i(h))=k\frac{h}{n}
\end{equation}
where $k$ is some coefficient of proportionality. Then one can rewrite the equation \ref{exprew1} as follows:
\begin{equation}\label{exprew2}
    \Extn(R_{total}(h)) = \sum_{i=1}^n k\cdot\frac{h}{n} = k\cdot h
\end{equation}
It means that the expected reward of coins issued to the miner in the whole network is proportional to his effort in the number of hashes that he generated during mining and independent of $n$. This is a great property since it implies that small mining farms with little bandwidth and storage and big mining pools can coexist in JaxNet with negligible penalties to their earnings. So, we have the same balance which one has in the network with the single blockchain: the rewards and influence of each participant is proportional to his hash rate.

One can argue, what is a benefit for a miner to operate many shards if he can mine what he wants in one of them? It is time to remember about transaction fees, which we do not divide by $n$. They satisfy the equation \ref{exprew1}. Therefore the more shards he is mining, the more transaction fees he competes for. This extra earning should at least cover his expenses for a faster internet connection, extra data processing and storage.

We can summarize the points above into the following statement:
\begin{proposition}
Described in the paper, the blockchain network is balanced in terms that every participant is rewarded proportionally to his effort in maintaining the network.
\end{proposition}

\subsection{Block Content} \label{BlockCont}

In this subsection we are going to discuss the block structure in JaxNet. The content of the blocks is listed in \cref{tableBCheader,tableBCbody,tableSCheader,tableSCbody}. The purpose of records \texttt{version}, \texttt{timestamp}, \texttt{bits} and \texttt{nonce} are the same as in Bitcoin\cite{Satoshi} and other blockchains. Block headers in the Beacon Chain include hashes of the previous block headers and roots of the Merkle trees of respective block bodies. Shard blocks have a similar design. However, in JaxNet blocks contain some extra data. Also, instead of previous block hashes in shard chains, we use  previous blocks commitments as discussed in the \cref{superlight}.

\begin{table}[ht!]
\centering
%\captionsetup{width=.9\columnwidth,font=footnotesize,
%  justification=centering}
  \caption{Beacon Chain Block Header}
  \label{tableBCheader}
\resizebox{0.9\columnwidth}{!}{%
\begin{tabularx}{\columnwidth}{lllX@{\extracolsep{\fill}}}
\toprule
Field Size & name & Data type & Comments\\
\midrule
4	& \texttt{version}	       & uint32\_t    & \\	
32	& \texttt{prev\_block}     & char[32]     & Hash of the previous block in the BC\\
32	& \texttt{Bmerkle\_root}   & char[32]     & Block body Merkle tree root\\	
32  & \texttt{MMmerkle\_root}  & char[32]     & Root of Merge-mining tree\\
4	& \texttt{timestamp}	   & uint32\_t    & \\	
4	& \texttt{bits}	           & uint32\_t	  & Target of BC\\
8	& \texttt{nonce}	       & uint64\_t    & \\
45  & \texttt{tree\_encoding}  & uint8[]    & Encoding of the Merge-mining tree\\
4   & \texttt{\# of shards}    & uint32\_t    & The number of shards in JaxNet and expansion flag\\
\hdashline
\multicolumn{4}{c}{Maximal Aggregate size is 165 bytes}\\
\bottomrule
\end{tabularx}}
\vspace{.2cm}
\end{table}
Besides standard records, the Beacon chain block header \cref{tableBCheader} includes the number of shards in the network, the encoding of shard merge-mining tree, and its root.
\begin{table}[ht!]
\centering
%\captionsetup{width=.9\columnwidth,font=footnotesize,
%  justification=centering}
  \caption{Beacon Chain Block Body}
  \label{tableBCbody}
\resizebox{0.9\columnwidth}{!}{%
\begin{tabularx}{\columnwidth}{lllX@{\extracolsep{\fill}}}
\toprule
Field Size & Name & Data type & Comments\\
\midrule
%3840 & tree encoding & char[3840]  & Merge-mining proof, largest possible size in bytes. 
32	& \texttt{txn\_count}	    & var\_int      & BC transaction count\\	
?	& \texttt{txns}             & tx[]          & BC transaction records\\
\hdashline
\multicolumn{4}{c}{Maximal Aggregate size is 64 Kbytes}\\
\bottomrule
\bottomrule
\end{tabularx}}
\vspace{.2cm}
\end{table}
The Beacon chain block body includes transactions and their count. The transaction part of the block body is limited to 24Kb.

\begin{table}[ht!]
\centering
%\captionsetup{width=.9\columnwidth,font=footnotesize,
%  justification=centering}
  \caption{Shard Chain Block Header}
  \label{tableSCheader}
\resizebox{0.9\columnwidth}{!}{%
\begin{tabularx}{0.9\columnwidth}{lllX@{\extracolsep{\fill}}}
\toprule
Field Size & Name & Data type & Comments\\
\midrule
%4	& \texttt{version}	       & uint32\_t     & \\
32	& \texttt{PBC}             & char[32]      & Commitment of previous shard blocks \\
32	& \texttt{Bmerkle\_root}   & char[32]      & Block body Merkle Root\\	
%4	& \texttt{timestamp}	   & uint32\_t     & \\	
4	& \texttt{bits}	           & uint32\_t	   & Target of SC\\
4   & \texttt{MM\_number}      & uint32\_t     & Merge-mining number\\
\hdashline
\multicolumn{4}{c}{Aggregate size is 72 bytes.}\\
\bottomrule
\end{tabularx}}
\vspace{.2cm}
\end{table}
The shard-chain block header \cref{tableSCheader} is rather standard. However, it includes merge-mining number \texttt{MM\_number} and doesn't contain a nonce.

\begin{table}[ht!]
\centering
%\captionsetup{width=.9\columnwidth,font=footnotesize,
%  justification=centering}
  \caption{Shard Chain Block Body}
  \label{tableSCbody}
\resizebox{0.9\columnwidth}{!}{%
\begin{tabularx}{0.95\columnwidth}{lllX@{\extracolsep{\fill}}}
\toprule
\makecell[l]{Field\\Size} & Name & \makecell[l]{Data \\ type} & Comments\\
\midrule
3840 & \texttt{MM\_Proof}         & char[3840]    & Merge-mining proof (largest possible size in bytes).\\
640  & \texttt{Shard\_Proof}      & char[640]     & Merkle Proof of the shard header block (largest possible size in bytes).\\
165  & \texttt{BC\_header}        & block         & Respective BC header block \\
32	 & \texttt{txn\_count}	      & var\_int      & SC transaction count\\
 
?	 & \texttt{txns}              & tx[]          & SC transaction records \\
\hdashline
\multicolumn{4}{c}{Aggregate size without txns is 4645bytes at maximum.}\\
\multicolumn{4}{c}{Transaction part is around 24Kbytes.}\\
\bottomrule
\end{tabularx}}
\vspace{.2cm}
\end{table}
Block body in shard chains includes a data record \texttt{BC\_header}, Merkle proof "\texttt{Shard\_Proof}" that this particular block was merge-mined according to the protocol and, finally, merge-mining proof "\texttt{MM\_Proof}" proves that \texttt{MM\_number} in the respective block header is valid. The transaction part of the block body is limited to 24Kb. Every transaction record includes shard IDs in order to prevent replay attacks on accounts that use the same private keys on different shards.

\begin{figure}[ht!]
	\begin{center}
		\resizebox{0.9\textwidth}{!}{
			\begin{tikzpicture}
\node[labelnode, minimum width=134pt, minimum height=136pt,
label={[black]90:{\scalebox{1.4}{SC Block Header $\#N$}}}] (SCBlockN) at (0,0){};
\node[BlockRecord] (target) at ($(SCBlockN)+(0,1.68)$) {\scalebox{1.3}{Target}};
\node[BlockRecord, outer color=yellow!20] (MMnumber) at ($(SCBlockN)+(0,0.5)$) {\scalebox{1.3}{MM number}};
\node[BlockRecord,outer color=green!20] (PrevBComm) at ($(SCBlockN)+(-1.1,-1.18) $) {\scalebox{1}{
\begin{tabular}{@{}l@{}}
Prev.\\ Blocks\\ Comm.
\end{tabular}
}};
\node[BlockRecord,outer color=green!20] (BRoot) at ($(SCBlockN)+(1.1,-1.18)$) {\scalebox{1}{
\begin{tabular}{@{}l@{}}
BBody\\ Merkle\\ Root
\end{tabular}
}};

\node[BlockRecord,outer color=yellow!20] (MMR) at ($(SCBlockN)+(-9,-1.5)$)
{\Large Merkle Mountain Range};

\node[BlockRecord,outer color=yellow!20] (ShardMerkle) at ($(SCBlockN)+(0,-5.7)$)
{\scalebox{1.5}{
\begin{tabular}{@{}l@{}}
Shard\\ Merkle \\ Proof
\end{tabular}
}};

\node[BBodylabel] (SCbody) at ($(SCBlockN)+(7,-5.7) $) {\scalebox{1}{
\begin{tabular}{@{}l@{}}
SC\\ Block \\ Body
\end{tabular}
}};

\node[BlockRecord,outer color=yellow!20] (MMproof) at ($(SCBlockN)+(7,0) $) {\scalebox{1.3}{
\begin{tabular}{@{}l@{}}
Merge\\ Mining\\ Proof
\end{tabular}
}};

\node[labelnode, minimum width=132pt, minimum height=136pt,
%label={[black]90:{\scalebox{1.2}{BC Block Header}}}
] (BCBlock) at ($(SCBlockN)+(-8,-5.8)$){};

%\node[BlockRecord, outer color=yellow!20] (Nshards) at ($(BCBlock)+(0,2)$) {\scalebox{1}{\# of shards}};
\node[BlockRecord, outer color=yellow!20] (Tree encoding) at ($(BCBlock)+(0,1.7)$) {\scalebox{1.1}{Tree encoding}};
\node[BlockRecord] (TimeStamp) at ($(BCBlock)+(-0.7,0.6)$) {\scalebox{1.1}{TimeStamp}};
\node[BlockRecord] (version) at ($(BCBlock)+(-1.1,-0.6)$) {\scalebox{1.1}{Version}};
\node[BlockRecord] (nonce) at ($(BCBlock)+(-1.1,-1.7)$) {\scalebox{1.1}{Nonce}};
\node[BlockRecord,outer color=green!20] (SRoot) at ($(BCBlock)+(+1.1,-1.2)$) {\scalebox{1}{
\begin{tabular}{@{}l@{}}
Shard\\ Merkle\\ Root
\end{tabular}
}};

\node[BlockRecord, outer color=gray!20] (Junk1) at ($(BCBlock)+(1.4,0.6)$) {\scalebox{0.9}{Junk}};
%\node[BlockRecord, outer color=gray!20] (Junk2) at ($(BCBlock)+(0,-1.7)$) {\scalebox{0.9}{Junk}};
%\node[BlockRecord, outer color=gray!20] (Junk3) at ($(BCBlock)+(1.4,-1.7)$) {\scalebox{0.9}{Junk}};

\foreach \x in {0,1,2,3,4}			
{\node[minichainnode] (BC\x) 
at ({-13+\x*2},1.9) {};}
\foreach \x[count=\xx from 1] in {0,1,2,3}
{\draw[myarrow3] (BC\xx) -- (BC\x);}

\draw[myarrow2] (PrevBComm.west) .. controls +(-2,0) and +(2,0)
.. (MMR.east);
\draw[myarrow2] (MMnumber.east) .. controls +(2,0) and +(-2,0)
.. (MMproof.west);
\draw[myarrow2] (BRoot.east) .. controls +(2,0) and +(-2,0)
.. (SCbody.west);

\draw[myarrow2] (ShardMerkle.north) .. controls +(0,1) and +(0,-1)
.. (SCBlockN.south);
\draw[myarrow2] (SRoot.east) .. controls +(1.5,0) and +(-1.5,0)
.. (ShardMerkle.west);

\foreach \x in {0,1,2,3,4}
{\draw[myarrow2] ($(MMR.north)+(-2+\x,0)$) .. controls +(0,1.5) and +(0,-1.5)
.. (BC\x.south);}
			\end{tikzpicture}
		}
		\vspace*{3pt}
		%\captionsetup{labelformat=empty}
		\caption{\textbf{Shard Chain block in JaxNet}} \label{SCbody}
    \end{center}
\end{figure}
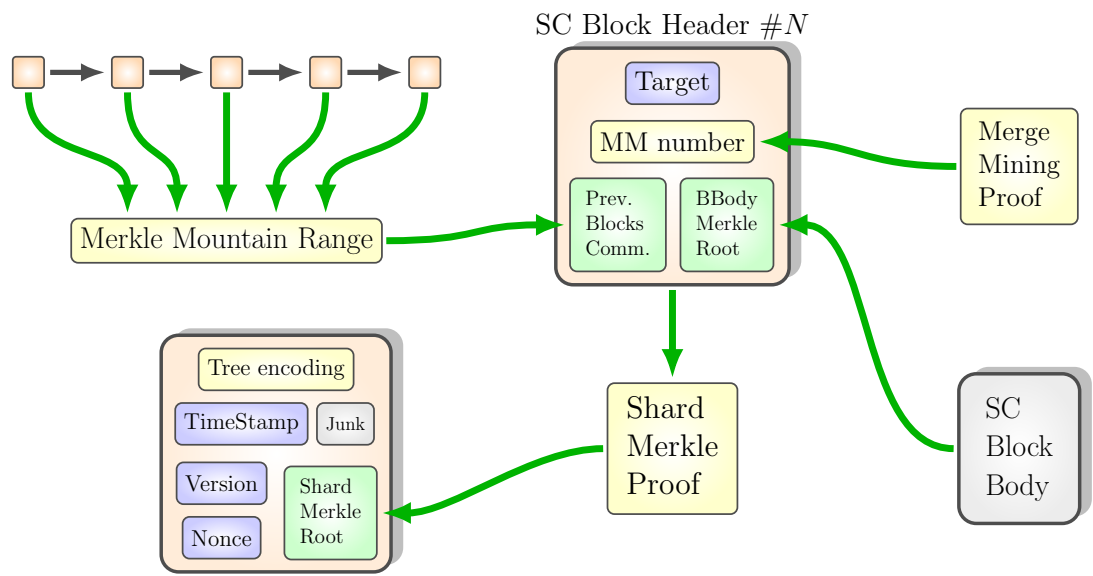

We need to highlight one important thing regarding \texttt{BC\_header} in SC block body. \texttt{BC\_header} is not necessarily a header of a given block on BC. It only has the same structure. This record will be referred as \texttt{BC\_header} container. It is important to avoid confusion here. In JaxNet there are no cross-chain links, which could be found in other blockchain proposals.  
\subsection{Mining scheme} \label{MMScheme}
\begin{figure}[ht!]
	\begin{center}
		\resizebox{0.7\textwidth}{!}{
			\begin{tikzpicture}
\node[labelnode,label={[black]90:{\scalebox{2}{BC Block Header $\#N$}}}]
(CMCBlockN) at (0,0) {};
\node[labelnode, draw=black!30, outer color=orange!5, label={[black]90:{\scalebox{2}{BC Block Header $\#N-1$}}}] (CMCBlockN-1) at (-10,0){};
\node[BlockRecord] (version) at ($(CMCBlockN)+(-1.9,2.2)$) {\scalebox{1.3}{Version}};
\node[BlockRecord] (timestamp) at ($(CMCBlockN)+(1.3,2.2)$) {\scalebox{1.3}{Timestamp}};
\node[BlockRecord] (target) at ($(CMCBlockN)+(-1.9,1.1)$) {\scalebox{1.3}{Target}};
\node[BlockRecord] (nonce) at ($(CMCBlockN)+(-2,0)$) {\scalebox{1.3}{Nonce}};
\node[BlockRecord, outer color=yellow!20] (Tree encoding) at ($(CMCBlockN)+(1.2,0)$) {\scalebox{1.3}{Tree encoding}};
\node[BlockRecord, outer color=yellow!20] (Nshards) at ($(CMCBlockN)+(1.2,1.1)$) {\scalebox{1.3}{\# of shards}};
\node[BlockRecord,outer color=green!20] (PrevBhash) at ($(CMCBlockN)+(-2.1,-1.7) $) {\scalebox{1}{
\begin{tabular}{@{}l@{}}
Prev.\\ Block\\ Hash
\end{tabular}
}};
\node[BlockRecord,outer color=green!20] (SRoot) at ($(CMCBlockN)+(+2.1,-1.7)$) {\scalebox{1}{
\begin{tabular}{@{}l@{}}
Shard\\ Merkle\\ Root
\end{tabular}
}};
\node[BlockRecord,outer color=green!20] (BRoot) at ($(CMCBlockN)+(-0.1,-1.7)$) {\scalebox{1}{
\begin{tabular}{@{}l@{}}
BBody\\ Merkle\\ Root
\end{tabular}
}};
\node[BBodylabel] (BCBody) at ($(CMCBlockN)+(-7.1,-7.7) $) {\scalebox{1}{
\begin{tabular}{@{}l@{}}
BC\\ Block\\ Body
\end{tabular}
}};
\node[BlockRecord,outer color=yellow!20] (ShardMerkle) at ($(CMCBlockN)+(0,-5.8)$)
{\Large Shard Merkle Tree};
\node[BBodylabel,outer color=orange!20,inner sep=4pt] (SCBody3) at ($(ShardMerkle)+(2,-2.8) $) {\scalebox{1}{
\begin{tabular}{@{}l@{}}
SC\\ Block\\ Header
\end{tabular}
}};
\node[BBodylabel,outer color=orange!20,inner sep=4pt] (SCBody2) at ($(ShardMerkle)+(0,-2.8) $) {\scalebox{1}{
\begin{tabular}{@{}l@{}}
SC\\ Block\\ Header
\end{tabular}
}};
\node[BBodylabel,outer color=orange!20,inner sep=4pt] (SCBody) at ($(ShardMerkle)+(-2,-2.8) $) {\scalebox{1}{
\begin{tabular}{@{}l@{}}
SC\\ Block\\ Header
\end{tabular}
}};
\draw[myarrow2] (PrevBhash.west) .. controls +(-3,0) and +(3,0)
.. (CMCBlockN-1.east);
\draw[myarrow2] (BRoot.south) .. controls +(0,-3) and +(0,3)
.. (BCBody.north);
\draw[myarrow2] (SRoot.south) .. controls +(0,-2) and +(0,2)
.. (ShardMerkle.north);
\draw[myarrow2] ($(SCBody.north)+(0,1.3)$) -- (SCBody.north);
\draw[myarrow2] ($(SCBody2.north)+(0,1.3)$) -- (SCBody2.north);
\draw[myarrow2] ($(SCBody3.north)+(0,1.3)$) -- (SCBody3.north);
			\end{tikzpicture}
		}
		\vspace*{3pt}
		%\captionsetup{labelformat=empty}
		\caption{\textbf{Merge Mining in JaxNet}} \label{MMinJaxNet}
    \end{center}
\end{figure}
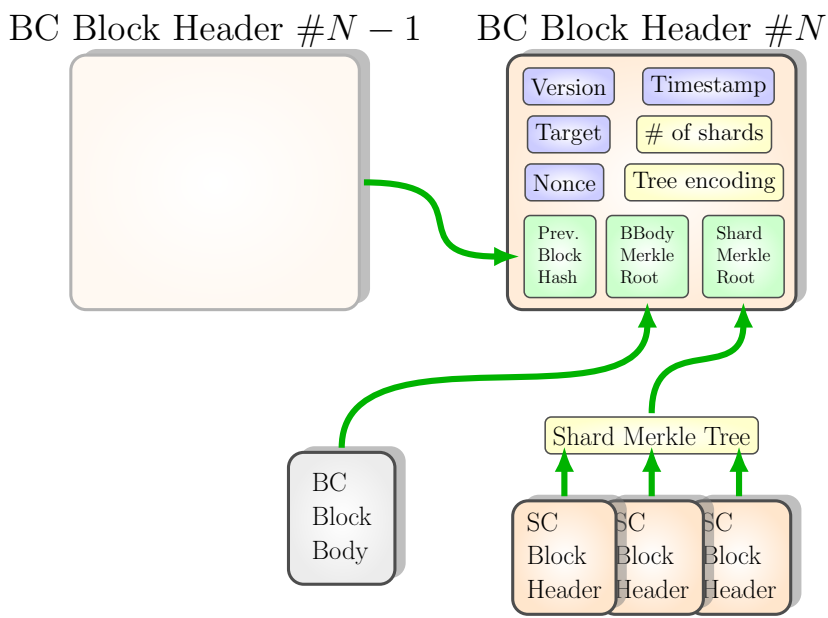

The mining scheme in JaxNet relies on merged-mining. The miner can choose the subset of shards, download the necessary information about them and the Beacon Chain, form valid blocks for those shards, put them into the Merkle tree as per \cref{MMinJaxNet} and start mining them simultaneously according to the following protocol:
\begin{enumerate}[1)]
    \item
Learn the total number of shards $N_S$ in JaxNet from the data in the Beacon Chain. Determine the depth $D_S$ of the Merkle tree. We call this tree Shard Merkle tree or Merged Mining Merkle Tree.
\begin{equation}
    D_S=\lceil log_2(N_S) \rceil
\end{equation}
%=\text{ceil}(log_2(N_S))$$
    \item
Choose the subset of shards for mining. Not every combination of shards is allowed. Check the subsection \ref{MMproof} on Merged-mining proofs. Also, the connection to nodes in each of those shards is required. Thus nodes that mine many shards need a good network connection.
    \item
Build shard blocks and put them into the Shard Merkle Tree (\cref{Merkle3}) as described in subsection \cref{MMproof}. Calculate the root of this Shard Merkle tree.
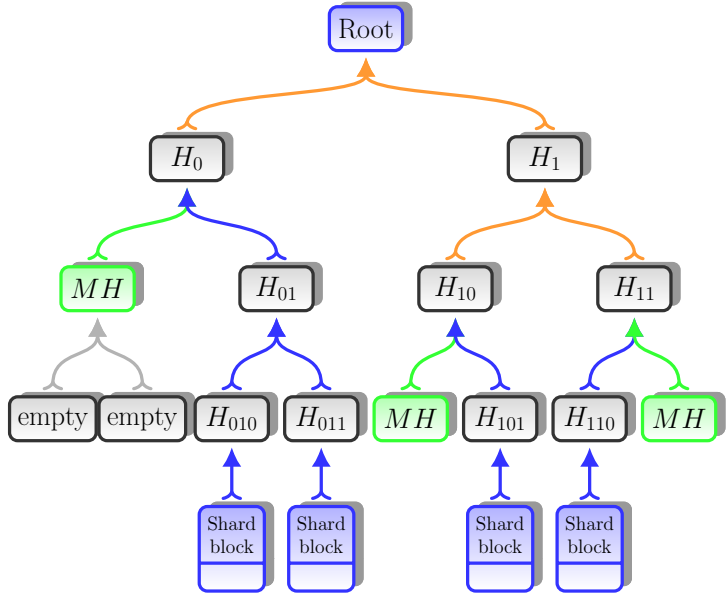
\begin{figure}[ht!]
	\begin{center}
		\resizebox{0.6\textwidth}{!}{
			\begin{tikzpicture}[
			%node distance=20mm,
			level distance=80pt,
			level 1/.style = {sibling distance=220pt},
			level 2/.style = {sibling distance=110pt},
			level 3/.style = {sibling distance=55pt},
			every node/.style=MerkleNode,
			every child/.style={mychild}
			]
\node[MerkleRoot] {Root}
    child[draw=orange!80] {node {$H_0$}
        child[draw=green!80] {node[MerkleProofNode] {$MH$}
            child[draw=gray!60] {node {empty}}
            child[draw=gray!60] {node {empty}}
        }
        child[draw=blue!80] {node {$H_{01}$}
            child[draw=blue!80] {node {$H_{010}$}
                child[draw=blue!80] {node[MerkleProofData,
                rectangle split parts=2]{Shard \\ block}}
            }
            child[draw=blue!80] {node {$H_{011}$}
                child[draw=blue!80] {node[MerkleProofData,
                rectangle split parts=2]{Shard \\ block}}
            }
        }
    }
    child[draw=orange!80] {node {$H_1$}
        child[draw=orange!80] {node {$H_{10}$}
            child[draw=green!80] {node[MerkleProofNode] {$MH$}}
            child[draw=blue!80] {node {$H_{101}$}
                child[draw=blue!80] {node[MerkleProofData,
                rectangle split parts=2]{Shard \\ block}}
            }
        }
        child[draw=orange!80] {node {$H_{11}$}
            child[draw=blue!80] {node {$H_{110}$}
                child[draw=blue!80] {node[MerkleProofData,
                rectangle split parts=2]{Shard \\ block}}
            }
            child[draw=green!80] {node[MerkleProofNode] {$MH$}}
        }
    };
			
%\node%[MerkleRoot]
%(b) 
%{Root}
%    child {\node%(bH0)
%    {$H_{0}$}}
%    child {\node%(bH1)
%    {$H_{1}$}
        %child 
            %{node(bH00){$H_{00}$}
            %child 
                %{node(bH000){$H_{000}$}
                %child {node[MerkleData]{DATA}}
%    };
			\end{tikzpicture}
		}
		\vspace*{3pt}
		%\captionsetup{labelformat=empty}
		\caption{\textbf{Merge Mining of shard blocks}} \label{Merkle3}
    \end{center}
\end{figure}
    \item
Put the root of the Shard Merkle tree into the \texttt{BC\_header} along with other valid data.
    \item
Merge-mine. During this process the miner generates a low hash of the \texttt{BC\_header}.
    \item
If the miner receives from the network or mines a new valid block for particular shard he should include it into his shard chain, rebuild respective shard block, update the respective leaf in the Shard Merkle Tree, recalculate intermediate hashes in it and the root, update the root in the \texttt{BC\_header}, continue mining.
    \item
If the miner successes in mining any SC block he broadcasts it to respective shard he connected to. If BC block is mined miner broadcasts it to every node he connected to. 
\end{enumerate}
For shard nodes it is not necessary to store \textit{BC\_headers} from SC block bodies and Merge Mining Proofs once respective SC block headers are rather deep in the shard chain.

Merge Mining scheme is displayed on the \cref{MMinJaxNet}.
\subsection{Merged mining proof} \label{MMproof}
\textit{Merged Mining Proof} or \textit{MMP} is the backbone and the key innovation of JaxNet protocol. According to JaxNet protocol, any miner can merge-mine multiple shards simultaneously. MMP is a tool that allows anyone within the particular shard to estimate how many shards were merge-mined. MMP is the proof that certain miner, who merge-mined the particular block, was not mining a particular subset of shards.

JaxNet protocol has the following steps and features which allow proof of merged mining to work. 
\subsubsection{Shard Merkle Tree}
%\begin{enumerate} [1)]
%    \item 
Nodes in JaxNet have a special agreement described in the protocol on how to perform merged mining. The set of shards in JaxNet are in one-to-one correspondence with the set of leaves of Shard Merkle Tree (\cref{Merkle3}): the first leaf from the left corresponds to the first shard, the second leaf to the second shard and so on. It is allowed to merge-mine the shard block only on the leaves of the Shard Merkle Tree which corresponds to this shard. Blocks which are merge-mined in the wrong place should be rejected during the block verification within the shard.

When the network decides to increase the network capacity, it may switch to higher Merkle Tree with more leaves and more capacity. The respective protocol is described in the \cref{ExpNet}.
\subsubsection{Magic hashes}
%    \item
When the miner calculates the root of the Shard Merkle Tree, he obeys the following rules. He starts from the lowest level of the Shard Merkle Tree, which consists of leaves. He calculates hashes in leaves that correspond to shards which have been chosen for merged mining. Each hash in the leaf is a hash of its respective SC block header. The rest of the leaves are left empty. Corresponding hashes of these empty leaves in the tree are considered to be \textit{undefined}. 

Then he calculates hashes level by level from the bottom to the top. The hash of the parent node is the hash of the concatenation of child nodes. If exactly one of the nodes is undefined, he substitutes it with the "magic hash" in Merged Mining tree. Magic hash consists of zero bits. If both child nodes are empty then the parent node is undefined too.

This construction establishes a simple way to prove that a certain shard was not mined. If corresponding to the shard leaf contains a magic hash or it is an ancestor of the node with such hash, then this shard was not mined by the miner when he generated the block with this Shard Merkle Tree. Therefore, in order to prove that at most $k$ shards were mined, it is sufficient to provide a valid Shard Merkle Tree with some amount of magic hashes in it.

For example, shards with binary indexes "000" and "001" on \cref{Merkle3} are ancestors of the magic hash in the node with index "00". Shards with indexes "100" and "111" correspond to leaves with magic hashes. These shards were not merge-mined when the block was generated.

Assume Shard Merkle Tree has a height $h$ and $h+1$ levels. Let's enumerate its levels from bottom to top so that leaves are on the level $0$ and root is on the level $h$. Let $m_i, i=\overline{0,n}$ is a number of magic hashes on the level $i$. Then this Shard Merkle Tree validates that
\begin{equation}
    \sum_{i=0}^{h} m_i \cdot 2^i
\end{equation}
shards were not merge-mined. Therefore at most 
\begin{equation}
    2^h-\sum_{i=0}^{h} m_i \cdot 2^i
\end{equation}
shards were merge-mined.

The construction similar to Shard Merkle Tree in Jaxnet was described by A.Zamyatin in his PhD thesis in subsection 2.2.1.\cite{zamyatin2016merged}

%    \item
\subsubsection{Merged Mining Proof}
When a Shard Merkle Tree root is calculated it is placed into the BC block header. Then the miner starts merged mining by changing the nonce in this BC block header and computing its mining hash which is described in the \cref{HashFunctions}. Assume he succeeds and gets a mining hash that is less than target of some SC block header. Then, he can compose the SC block described in the \cref{BlockCont} and display it on the \cref{SCbody}. 

Since a Shard Merkle Tree protects the data integrity, the miner has a proof of how merged mining has been executed. A Shard Merkle Proof is a merkle proof that the SC block header was mined in the correct place. \texttt{MM\_number} is the miner's claim about how many shards he was mining. In the \cref{JaxnetRew} it is discussed how \texttt{MM\_number} is used to set a proper reward for the SC block. \texttt{Merged Mining Proof} and Shard Merkle Tree encoding is his proof that \texttt{MM\_number} is valid. 

%\end{enumerate}
\subsubsection{Orange subtree and encoding}
%Let us consider the subtree of the Shard Merkle Tree with the same root, and which leaves are parents of the nodes with magic hashes.
According to our construction of Shard Merkle Tree, magic hashes are placed in those nodes of Shard Merkle Tree which doesn't have any shard block candidates as successors in Shard Merkle Tree and whose parent, if it exists, doesn't have this property. Let's recall these nodes as "magic nodes". 

We define \textit{regular nodes} in the similar way. We call a node a regular node if all leaves among its successors were used as mining slots for block candidates, and whose parent doesn't have this property if it exists.

Also there is a third category of nodes. We call ancestors of magic nodes and regular nodes as orange nodes. Orange nodes form a subtree of Shard Merkle Tree with the same root.

Let us call this subtree as "orange subtree" within this paper. For example, on \cref{Merkle3} edges of this subtree are colored in orange. Its leaves are nodes with labels "$H_0$", "$H_{10}$" and "$H_{11}$". 

Notice that leaves of the orange subtree with $n$ nodes has $n+1$ children. The set of these children coincide with the union of the set of magic nodes and the set of regular nodes. If we extend orange subtree with this nodes we obtain a full binary tree. Let's call this tree a \textit{full orange subtree}.

%If there are no parents of magic hashes in the Shard Merkle Tree then the respective orange subtree is considered to be empty. 
The full orange subtree always contain at least one node. However, orange subtree could be empty. This is possible only either if the miner doesn't mine any shard or if he mines all shards and the shard count is the power of two. 

In the first scenario the root of Shard Merkle Tree contains a magic hash. This root is included in the respective \texttt{BC\_container}. So any other miner can determine this case by looking at the content of \texttt{BC\_container}. However, in this case it's not possible to mine any shard block through merge-mining. Thus any shard block candidate with such \texttt{BC\_container} should be rejected.

In the second scenario if the shard count is the power of $2$m there are no need in Merged Mining Proof. Other miners assume that the miner was mining all shards. Merged Mining Proof is considered to be empty.

Let us assume that orange subtree is non-empty.

Let's denote the set of nodes of orange subtree as $I$, the set of nodes with magic hashes as $M$ and the set of regular nodes as $P$.

In JaxNet, Merged Mining Proof is the list of hashes that correspond to nodes in $P$ together with the encoding of the orange subtree. If the full orange subtree has $k+1$ leaves then
\begin{equation}
    |I|+1=|M|+|P|=k+1.
\end{equation}

%Let us denote the set of leaves of the orange subtree as $L$, the set of nodes with magic hashes as $M$. %We see that each node in $L$ has two children: one of them is in $M$ and the second one is some regular hash. Let's denote the set of these second children as $P$. (We neglect the probability that this regular hash is the magic hash). 
%In JaxNet, Merged Mining Proof is the list of hashes that correspond to nodes in $P$ together with the encoding of the orange subtree. If an orange subtree has $k+1$ leaves then
%\begin{equation}
%    |L|=|M|=|P|=k+1.
%\end{equation}
Since orange subtree is not empty, Merge Mining proof will contain at most $k$ hashes. We are now left to determine what is encoding in the orange subtree. This encoding consists of two parts. One of them is the structure of full orange subtree as a full binary tree. This binary tree encoding has $2k-1$ bits. It could be found in the \cref{TreeEnc}. 

The second part of the encoding is the position of regular and magic hashes. Each leaf of orange subtree has two children: one is a magic hash and second is regular hash. However, it's not known which hash is on the left and which one is on the right. One bit per leaf is needed to encode this position. In total, $k+1$ bits are required.

Therefore, orange subtree encoding has $3k$ bits. 

It is not mandatory to take an orange subtree with a certain number of leaves. This number is variable. If orange subtree has less than $k$ leaves then encoding will be shorter. Limitations on $k$ are discussed in the \cref{TreeEnc}.
%The number of binary trees with $k+1$ leaves is known to be the Catalan number $C_k$. The number of possible orange subtrees is bounded from above by the sum of the first $k$ Catalan numbers. Binary trees with at most $k+1$ leaves could be encoded with $2k$ bits. Another $k+1$ bits are required to save the position of nodes in $P$ with respect to their parents in $L$.
\subsection{Expanding the Network} \label{ExpNet}
The key feature of this scalable solution is the ability to effectively address the problem at any scale. However, in each particular case some choice of parameters may work better than another. It is easy to choose parameters if there is a central authority that knows well the current state of the system. The question is how to set these parameters in the distributed network.

The key parameter in JaxNet is the number of shards $N$. When the number of users in the system and transaction count is limited, the most efficient solution is to maintain only a few shards. One chain solutions could perform even better. Besides efficiency, having too many shards can trigger a security risk. If there are too many shards then there will be little incentive to merge mine them for powerful nodes. Thus, the hash rate in many of them could plummet. So the selection of parameter $N$ is a rather big responsibility.

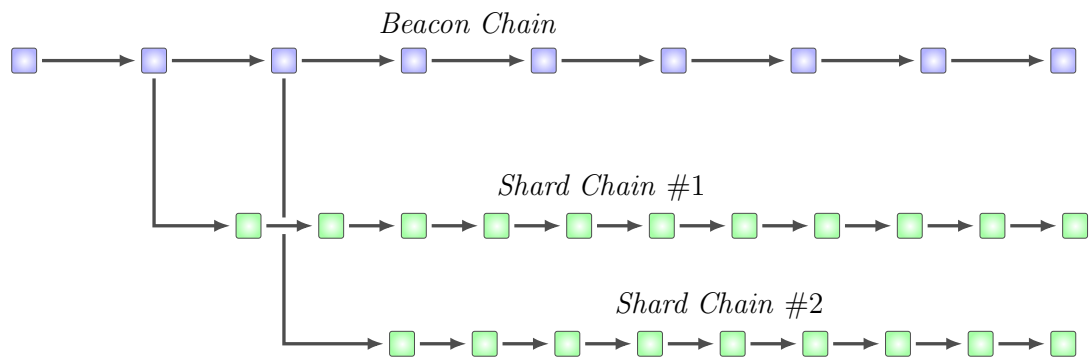
\begin{figure}[ht!]
	\begin{center}
		\resizebox{0.9\textwidth}{!}{
			\begin{tikzpicture}
%[start chain={going right},
%every node/.style={on chain},
%every join/.style={myarrow},]
\foreach \x in {0,1,2,3,4,5,6,7,8}			
{\node[chainnode,outer color=blue!30] (BC\x) 
at ({\x*5.5},0) {};}
\foreach \x[count=\xx from 1] in {0,1,2,3,4,5,6,7}
{\draw[myarrow] (BC\xx) -- (BC\x);}

\foreach \x in {0,1,2,3,4,5,6,7,8}			
{\node[chainnode] (SCb\x) 
at ({16+\x*3.5},-12) {};}
\draw[myarrow] (SCb0.west) -| (BC2.south); 
\foreach \x[count=\xx from 1] in {0,1,2,3,4,5,6,7}
{\draw[myarrow] 
(SCb\xx) -- (SCb\x);}

%\pgfsetfillopacity{1}
%\begin{pgftransparencygroup}
\foreach \x in {0,1,2,3,4,5,6,7,8,9,10}			
{\node[chainnode] (SCa\x) 
at ({9.5+\x*3.5},-7) {};}
\draw[myarrow] (SCa0.west) -| (BC1.south);
\foreach \x[count=\xx from 1] in {0,1,2,3,4,5,6,7,8,9}
{\draw[myarrow, draw=white, line width=20pt] (SCa\xx) -- (SCa\x);
\draw[myarrow] (SCa\xx) -- (SCa\x);}
%\end{pgftransparencygroup}

\node[text width=10cm] at (20,1.6) {\scalebox{3}{\textit{Beacon Chain}}};
\node[text width=10cm] at (25,-5.4) {\scalebox{3}{\textit{Shard Chain} \#1}};
\node[text width=10cm] at (30,-10.4) {\scalebox{3}{\textit{Shard Chain} \#2}};
			\end{tikzpicture}
		}
		\vspace*{3pt}
		%\captionsetup{labelformat=empty}
		\caption{\textbf{Chains in JaxNet}} \label{MCbody}
    \end{center}
\end{figure}

In JaxNet, we set a simple consensus on the value based on the hash rate. Block headers on the Beacon Chain contain a specific field. The last bit in this field is the flag. The rest of the bits are the binary representation of the number of shards in the system. 

A flag with value '1' signals that miner who mined this block votes to increase the number of shards in the system and '0' signals about the opposite. On each mining round miners count the number of '1' in previous blocks on the Beacon chain. 
If all conditions below are satisfied:
\begin{enumerate}[a)]
    \item
number of ones in the previous $1024$ blocks on BC is greater than $768$ 
    \item
in the previous $1024$ blocks on BC the number of shards was constant.
\end{enumerate}
then on the next block shard number $N$ have to be increased by
\begin{equation}
    dN = max \left( 1,2^{\lceil \log_2(N) \rceil-9} \right)
\end{equation}
Increasing the number of shards in the block number $n$ triggers the following events. Starting from block $n+10$ miners can merge mine shards $N+1$,..., $N+dN$. Genesis block on those shard chain has the following content. The hash of the previous block in it is the hash of the $(N+9)$th block on the Beacon Chain. The target of the genesis block is the target of the $(N+9)$th block on the Beacon Chain multiplied by $80$.

If the $N$ is the power of $2$ then the Shard Merkle tree is full. Thus after $N$-th block miners merge mine on the Shard Merkle Tree which is one level higher.

Blocks on any chain with an incorrect number of shards or with a wrong height of the Shard Merkle tree should be rejected.

\subsubsection{Alternative protocol for network expansion}
In the alternative protocol for the network expansion, there is an extra step. Every shard could be marked with a flag "full" according to predefined criteria of fullness. This criteria is based on the amount of space in shard blocks that was occupied by transactions within the target interval on the chain. 

Miners in every BC container include information about some subset of shards which are currently, or were flagged, as "full". Then this data is used to reach consensus on the question of whether a new shard should be created. SC blocks could be rejected based on incorrect flags in the associated BC container. So theoretically, the malicious actor who attempts to place invalid flags into the BC container, could get a penalty since some of his blocks in shards will be rejected.

However, this approach doesn't bring improvement against shard expansion protocol. In such an attack the malicious actor attempts to force the creation of many new unnecessary shards. The main part of the attacker's expense is acquisition of mining facilities. 

The alternative protocol doesn't increase the amount of mining rigs that are required to perform the attack. Also, it makes possible new attack vectors since the attacker gets an option to spam some shards with transactions in order to persuade the community that many shards are full and there is a need for new shards. The cost of an attack that utilizes multiple attack vectors and sophisticated strategy often will be cheaper than a straightforward one based on hash rate supremacy.
\subsection{Block Verification}
Block verification on the Beacon chain for full nodes follows the common standards for single-chain blockchains, plus few extra steps. It should be verified that the number of shards is valid and calculated according to the protocol. Also, the tree encoding size should not exceed the size limit. 

On shard chains verification is more tricky.
\begin{enumerate} [1)]
    \item
First, check that all data in SC block header is available, has prescribed types and size limits. Check version and timestamps.
    \item
Check that the hash of the \texttt{BC\_header} is below the target of the shard.
    \item
Verify that \textit{MM\_number} in SC header, shard count in block's \texttt{BC\_header} container and current shard count determined by expansion protocol satisfy the inequality:
\begin{equation}
\textit{MM\_number} \le \text{block's shard count} \le \text{current shard count}
\end{equation}
    \item
Take \textit{Shard\_Proof} from the SC block body and verify that this body corresponds to the SC block header.
    \item
Reconstruct the orange subtree (see subsection \cref{MMproof}) from the encoding stored in \texttt{BC\_header}.(check \cref{TreeEnc}) Take \textit{MM\_Proof} and verify that Merge Mining Proof is valid. In particular check that \texttt{MM\_number} is proved to be valid.
    \item
Check whether shard transactions from the SC block body are valid. Note that in JaxNet there will be set a minimal transaction fee.
\end{enumerate}

\subsection{Super light client} \label{superlight}
In the design of the Bitcoin network Satoshi Nakamoto introduced the construction of \textit{light client}. It provides devices with limited resources a way to verify transactions by getting proofs from full nodes using \textit{Simplified Payment Verification} (SPV).\cite{Satoshi} However, to make any SPV possible the light client has to determine the valid chain. Hence, it is necessary to download all headers in the chain. In the case of Bitcoin the aggregate size of block headers in 2019 is around 50Mb. However, for other cryptocurrencies this number could be significantly higher. For instance, block interval times are 40 times less and their average size is 6-7 times larger than in Bitcoin. Therefore, even light client with SPV becomes demanding.\cite{bunz2019flyclient}

In the past decade the blockchain community has proposed few approaches to address this problem by constructing so-called \textit{super-light client}. These approaches endeavor to make use of so-called \textit{reference blocks} or \textit{super blocks}. It's worthwhile to mention the work of Kiayias \textit{et al.} on \textit{proofs of proof of work} (PoPoW)\cite{kiayias2016proofs} and \textit{non-interactive PoPoW} (NiPoPoW)\cite{kiayias2017non}. However, this design has multiple drawbacks. First, it does not work properly if block difficulty is not fixed. Second, some bribery attacks were proposed. Third, there are some concerns about efficiency. Proofs, that a certain transaction was included in a certain block requires the download of the subchain of block headers.\cite{bunz2019flyclient}

Recently B{\"u}nz \textit{et al.} proposed the design of FlyClient which aims to resolve the problem for blockchains with variable difficulty.\cite{bunz2019flyclient} Also, this solution is a significant step forward in terms of efficiency. It is based on \textit{Merkle Mountain Range} commitments and Fiat-Shamir heuristic\cite{fiat1986prove}. The author of FlyClient proposes a way to deploy it in existing blockchains by including Merkle Mountain Range tree roots in block headers.

JaxNet incorporates the idea of a FlyClient in its design. In particular, we use Merkle Mountain Range tree roots instead of previous block hashes. There is no sense to keep both of them in the block header. We recall Merkle Mountain Range tree roots in JaxNet as \textit{previous blocks commitments} (PBC) since they play the same role as previous block hashes in other blockchains. Also, Merkle Mountain Range in JaxNet follows the design of \textit{FlyClient Under Variable Difficulty} described in the same paper\cite{bunz2019flyclient}. However, there are few differences.

\begin{enumerate}
    \item 
Leaves of Merkle Mountain Ranges are hashes of concatenation of SC headers and hashes of \texttt{BC\_header} containers associated with them. The reason for this decision is described in the \cref{SecMod}.
    \item
In the original design\cite{bunz2019flyclient} of FlyClient nodes of the Merkle Mountain Range contain the data about timestamps and difficulty so that the super-light client can verify correctness of difficulty transitions based on timestamps. It was done in order to prevent the \textit{difficulty raising attack}. However, the paper contains an inaccurate argument regarding this setting. It is discussed in the \cref{FlyBugs}.

In contrast to the original design the super-light client in Jaxnet doesn't verify difficulty transitions in the chain. Still there is a verification that the particular chain is the heaviest one. If honest nodes control the majority of the computational resources, we can assume that they maintain the heaviest chain and all difficulty transitions in it are valid. 

Every Merkle Mountain Range node in JaxNet does not contain the data regarding timestamps. However, it includes the record "subchain weight". Subchain weight of the leaf is the difficulty of the respective block in the chain. Subchain weight of the regular node is the sum of weights of its children. Therefore this record contains the aggregate weight of blocks below it. This design makes possible the construction of the proof described in the original paper that is required for the verification of the chain weight. In contrast, in the original design there were two records: one for the chain weight at the beginning of the subchain and one for the chain weight at the end of the subchain. So the subchain weight was calculated as the difference between these two records.
    \item
Sampling
\end{enumerate}

A detailed description of FlyClient implementation and performance in JaxNet goes beyond the scope of this paper. 
\subsection{Timestamps and difficulty} \label{DiffandTimeS}
In the Bitcoin network, \cite{Satoshi} and similar solutions, it is a common practice to regulate block times, block difficulties, and the coin issuance rate, through some mechanism based on timestamps. JaxNet has a list of such rules as well. However, these rules are slightly different. We are going to describe them in this section. First, we share the motivation of these rules. Desired goals are as follows: 
\begin{enumerate}[a)]
    \item
    Keep the block time on the Beacon Chain around 600s.
    \item
    Keep the block time on every Shard Chain around 15s.
    \item
    Keep the difference between the creation time of the $(N+1)$-th block on the Beacon Chain and the creation time of the first block around $600\cdot N$ seconds.
    \item
    Keep the difference between the creation time of the $(N+1)$-th block on every Shard Chain and the creation time of the first block around $15\cdot N$ seconds.
\end{enumerate}
We set a priority for goals "c" and "d". This setting differs from the majority of other cryptocurrencies in which rules are designed to keep a fixed block time and certain coin issuance. In JaxNet, the "heaviest chain rule" is more important than the "longest chain rule" in the consensus. Besides, chain length is a useful tool to set a "soft synchronization" between chains. Synchronization in a distributed network is a rather hard task. However, in certain case of the rather short fork chain length is a tool which can do the job. Also, it is cheaper from the viewpoint of both communication and computation.        

\subsubsection{Difficulty Adjustment Algorithm}
The difficulty for the first epoch on the BC is set to be $D_0$. The difficulty for the first epoch of the SC equals half of the difficulty of respective genesis block on the BC for this SC. 

Block difficulty is adjusted every epoch. For the SC an epoch is 
\begin{equation}
    N_{SC}= 4 \cdot 60 \cdot 24=5760
\end{equation}
blocks. For the BC it is
\begin{equation}
    N_{BC} =2^{11} = 2048 \simeq 2016 = 2\cdot7\cdot 24\cdot 6
\end{equation}
blocks. Similarly to Nakamoto's design, we allow a time window for accepting timestamps. 

Let's set some notations. Let 
%\begin{itemize}
%    \item
    %$H_m$ denotes the $m$-th block header either on BC or on SC.
    
%    \item
    %$D(H_m)$ is a difficulty of the $m$-th block.
    
%    \item
    $D_0$ is a minimal allowed difficulty in the network.
    
%    \item
    $T_{BC}=600$ is a desired block time of BC.
    
%    \item
    $T_{SC}=15$ is a desired block time of SC.
    
    %$T_{prev}$ is a target block time for the previous epoch.
    
    $T_{next}$ is a target block time for the next epoch.
    
    $D_{prev}$ is the target difficulty for the previous epoch.
    
    $D_{next}$ is the target difficulty for the next epoch.
    
    %$\tau(H)$ is a timestamp of the block header $H$.
    %$\tau_3'$ is a median of timestamps of first $5$ blocks in the epoch.
    %$\tau_{n-2}'$ is a median of timestamps of last $5$ blocks in the epoch.
%\end{itemize}

The Difficulty Adjustment Algorithm in JaxNet works as follows. Assume the previous epoch has ended on the block $n$ and $\tau_n$ is its timestamp.
\begin{enumerate}[1)]
    \item 
    First, we calculate how the actual timestamp of block $n$ differs from the desired timestamp of this block. We start from the genesis block of the BC. This block is number $0$ in the chain and the timestamp $t_0=0$. Then, given the desired block time for the BC is $T_{BC}=600$, the desired timestamp for the block number $n$ in seconds is $t_n=600n$. 
    
    For the SC the genesis block is some block \#$m$ on the Beacon Chain. It has the desired timestamp $600m$. For the block \#$n$ on the SC the desired timestamp in seconds is defined as $t_n=600m+15n$.
    
    On the first step the node calculates the difference:
    \begin{equation}
        d_n = \tau_n-t_n
    \end{equation}
    \item
    Motivation of the second step is to estimate the average hash rate during the previous epoch.
    
    Let $\tilde{\tau}_1$ is a median timestamp of first $5$ blocks in the previous epoch and $\tilde{\tau}_2$ is a median timestamp of the last 5 blocks in that epoch.
    
    Then the average hash rate $AHR$ during that epoch could be estimated as
    \begin{equation} \label{eq:AHR}
    AHR=\frac{(N-5) \cdot D_{prev}}{ (\tilde{\tau}_2-\tilde{\tau}_1)}
    \end{equation}
    where $N$ is a number of blocks in the epoch.
    \item
    The goal of the third step is to adjust the difficulty for the next epoch in such a way that the expected value of the timestamp of the last block in it coincide with its desired value.
    \begin{equation} \label{eq:newBT}
        \frac{D_{next}}{AHR} = \frac{d_n+N\cdot T}{N}
    \end{equation}
    where $T=T_{BC}$ for BC and $T=T_{SC}$ for SC.
    
    Formally, the fraction in the left-hand-side of \cref{eq:newBT} is the desired expected block time $T_{next}$ for the next epoch. In order to avoid too long or too short values of this fraction let us agree to adjust it as follows:
    \begin{equation} \label{eq:adjnewBT}
        T_{next}=\frac{D_{next}}{AHR}=
        \begin{cases}
        \frac{d_n+N\cdot T}{N} &
        \text{if } 0.8T<\frac{d_n+N\cdot T}{N}<1.2T\\
        0.8T &
        \text{if } 0.8T\ge\frac{d_n+N\cdot T}{N}\\
        1.2T &
        \text{if } \frac{d_n+N\cdot T}{N}\ge 1.2T
        \end{cases}
    \end{equation}
    The difficulty $D_{next}$ for the next epoch is calculated from \cref{eq:AHR} and \cref{eq:adjnewBT}. 
\end{enumerate}

\subsubsection{Timestamp Window}
Clock synchronization between computer systems is not an easy task. Existing solutions such as Network Time Protocol (NTP) \cite{mills2010network} could cause errors. Also, due to various reasons, a miner's node could go offline for a short time and lose synchronization. 

Therefore, blockchain networks often set some \textit{Timestamp Window} within which the timestamp is considered to be valid. In the original Bitcoin design, Time Window was set to be $2$ hours. Recent research \cite{TimeTrS} shows that over the last decade mining became more professional and timestamp errors seldom exceed $1$ minute. This is a reason why in JaxNet, the Time Window on SC is reduced to be 6 minutes. The time window on BC remains at $120$ minutes.

Literally, timestamp control rules \cite{Timestamps} for BC are the same as in Bitcoin. Timestamp $T(H_n)$ of block header $H_n$ is rejected if anything below is true:
\begin{itemize}
    \item (\textit{Median Past Time (MPT) Rule})
    \begin{equation}
        T(H_n) \le \text{median of the previous 11 blocks}
    \end{equation}
    \item (\textit{Future Block Time Rule})
    \begin{equation}
        T(H_n) \ge \text{median time of the peers' nodes}+2h
    \end{equation}     
\end{itemize}

For SC there are following rejection rules:
\begin{itemize}
    \item %(\textit{Median Past Time (MPT) Rule})
    \begin{equation}
        T(H_n) \le \text{median of the previous 23 blocks}
    \end{equation}
    \item %(\textit{Future Block Time Rule})
    \begin{equation}
        T(H_n) \ge \text{median time of the peers' nodes}+6m
    \end{equation}     
\end{itemize}
It is important to remember that a timestamp rejected now could become valid in the future. 
\subsection{Hashing algorithms} \label{HashFunctions}
As one may notice, the design of JaxNet heavily uses cryptographic hash functions, or simply hashes. In this subsection we will specify what particular hash functions are used.

The BLAKE2s hashing function is often used in Merkle trees, block header hashes, etc. The exception is the hash of the Beacon chain block header. In this case, we use a special hash function described below in this section. In the design of JaxNet, this hash works as a mining hash. So, in the Proof of Work challenge, miners are searching the nonce such that mining hash of the respective BC block header is below the target. Moreover, merged mining of shard blocks requires the finding of a low BC block header hash as well, as discussed in the \cref{MMproof}.

The motivation is the following. For Merkle tree constructions we need a fast, efficient, hash algorithm. JaxNet uses BLAKE2s that is well-known and poses desirable properties. However, the authors of this paper believe that the mining hash function should pose one extra property called \textit{ASIC-resistance}. Many researchers say with confidence that complete ASIC-resistance is unachievable. Nevertheless, some hash functions are known to be more ASIC-resistant than others.

The latest revision of the Ethash algorithm \cite{Ethash} is known for its good ASIC-resistance. However, it appears that it has some incompatibility issues with FlyClient discussed in the \cref{superlight}. JaxNet uses a modified version of Ethash as its mining algorithm. However, the description of this hash function goes beyond the scope of this paper.
\subsection{Cross-shard transactions}
Exchanging coins between nodes in different shards is a big problem.\cite{zamyatin2019sok,wang2019sok,sonnino2019replay} One possible solution is to use a trusted third party. Although, some other approaches were proposed.\cite{skidanov2019nightshade,javad2019sharper}
%,martino2018chainweb}

In short, the problem is as follows. Assume Alice has a non-empty wallet in a shard and wants to send part of her funds in it to Bob. Assume Bob doesn't have a wallet and his JaxNet client has no information about this shard. In JaxNet shard chains are independent of each other. Therefore, there is no way to directly transfer coins from one shard to another. However, there are some tools and features which resolve this problem.

%Jaxnet has a valuable feature for settling reliable cross-shard transactions.
\begin{enumerate}[1)]
    \item 
First, coins in different shards have almost the same value and could be exchanged one-to-one.
    \item
Second, Jaxnet has a Super Light Client discussed in the \cref{superlight}. A party can easily connect to the shard by downloading few MB of data. Therefore, Bob can open a wallet in this shard and synchronize his client with a shard chain. 

Then he can check whether the transaction from Alice was included into the block on SC. However, if Bob's client got synchronized with many chains it could become heavy. Thus, there is a need for a third party that can make optimizations on request in exchange for  fees.
    \item
In JaxNet, hubs take the role of a third party for some transactions. Hubs assist in moving funds between shards. Any user can send coins from his wallet in one shard and receive coins from the hub in another shard. No permission is needed to set a hub on a node. Users are free to choose the hub. However, there is a need for trust between users and the hub.  
\end{enumerate}
%Nevertheless 

%Another possible solution is using Super Light Client discussed in the \cref{superlight}. As of May 2020, there is no state-of-the-art solution to this problem. JaxNet is friendly to smart contracts and other possible approaches, and doesn't have a unique or "recommended" solution to this problem. We leave it to competition as long as trust is maintained. 
\subsection{Security model} \label{SecMod}
A blockchain network is a rather complex system which could be attacked from multiple vectors.\cite{saad2019exploring} We will discuss those which are specific to blockchain designs and most relevant for JaxNet security. 
\subsubsection{51\% attacks}
One of the main threats of blockchain sharding schemes is \textit{single-shard takeover attacks}\cite{EthSharding}. They are the sort of $51\%$ attacks performed within the shard. JaxNet attains some level of security against them with merge mining.

Even though an attacker can perform a successful attack after aggregating around $25\%$ of the hash rate, this $25\%$ is not fixed. We assume that new shards are established when a coalition of nodes is able and willing to sustain it. It is assumed that a new shard is opened responsibly, so that there is an appropriate hash rate to protect it from single-shard takeover attacks. 

It is assumed that a coalition of nodes as mentioned above will supply the required hash rate. Of course, at some point the hash rate in the shard may drop below this $25\%$ threshold. This is the price we have to pay for scalability. 

One may argue that this price is unacceptable. Nonetheless, the percentage of the network hash rate is not always the best measure of security. As an example one can compare Bitcoin, and a given unknown altcoin, based on the same hashing algorithm. One can easily conduct double-spending attacks on unknown altcoin, however it will be very hard and expensive to repeat that trick with Bitcoin.\cite{grunspan2019profitability} This situation may change in the future, however in November 2019 this fact can be accepted with confidence.

As another approach to measure the security of the network one may consider \textit{security budget} and \textit{security factor}.\cite{BTCsecbud1,BTCsecbud2} Annual security budget is an amount of money paid in coins to miners in block rewards and transaction fees. Roughly speaking, it is a number of coins multiplied by their price in US dollars, or in another asset. This money should cover their expenses on mining hardware and electricity.

Assuming that mining hardware, on average, is effectively working throughout the year, we can suggest that this parameter gives an estimate of how much money is invested in mining hardware by honest nodes. In order to perform a successful attack a malicious actor should make a comparable investment. Purchasing or renting a large amount of hardware is considered to be a hard task, discouraging said illicit investment. 

Obviously, if potential profit exceeds expenses, security budgets will not stop an attacker. However, the security budget provides understanding of what amount of money is safe to transfer through the blockchain.

For the large blockchain network, it is good to have a high security budget in order to protect large transfers. The security factor is a fraction which characterises this ratio:
\begin{equation}
    \text{security factor}=\frac{\text{security budget}}{\text{capitalization}}
\end{equation}

In order to maintain the system securely we should keep the parameters mentioned above on a high level. Since in JaxNet, the cost of an attack denominated in percents of the hash rate is less than on single-chain blockchains, it is important to have a higher security budget and security factor. This does not mean JaxNet is less secure. It means we pay, for the same level of security, more than is required for the less capable, more casual blockchain. Such is the nature of secure global transactions.

%\subsubsection{Data availability attacks and DDOS}
%JaxNet use security mechanisms utilized in Etherium as well as its open source code and blockchain constructor.\cite{wood2014ethereum}

\subsubsection{Timestamp cheating}
There are known attacks on blockchain networks based on timestamp manipulation. If malicious nodes control around 50\% of the network hash rate they may try to increase the block rate in the network. In networks similar to Bitcoin such actions could be rather profitable.\cite{davidson2020profitability} %However, timestamp cheating is easy to spot. There

In JaxNet it is possible to perform this attack if one poses a significant hash rate. However, the benefit of such an attack is lessened since the reward is almost independent from the block rate. Implementation of inflation management from the \cref{InfMan} could increase vulnerability. However, even in this case any attack will require significant computational power and little expected profit. Block rate in affected chains will recover to the normal state after few difficulty adjustments.

\subsubsection{Attacks on Proof of Merged Mining mechanism}
Malicious actors can attempt to attack Proof of Merge Mining Mechanism in JaxNet. Since the hash function validates the integrity of the content, an attacker can not replace data fields after \texttt{BC\_header} container with low hash was generated. He may try to merge-mine the SC block in multiple places. However, each shard has a prescribed position in Shard Merkle tree. If the attacker try put SC header on different level his attack will fail. Respective \texttt{BC\_header} container has a record of how many shards are in the network. Based on this record only one level in Shard Merkle tree will be valid.

Another attack vector is generating \texttt{BC\_header} container for some SC header within the chain. It could be a problem for the Super Light Client described in the \cref{superlight}. The attacker can generate some chain of SC headers, generate a heavy-chain proof with Fiat-Shamir heuristics, get a list of important SC headers and only then validate them by generating \texttt{BC\_header} containers for them. In order to address this issue Merkle Mountain Range commitment in JaxNet includes commitment of previous \texttt{BC\_header} containers.

\subsubsection{Attacks on shard expansion mechanism}
Malicious actors may lobby to open too many shards in the network so that it will not be able protect all of them. It is assumed that honest miners will not vote for opening new shards if they are not going to mine them. Therefore opening extra shard against the will of majority will require huge expenses on maintaining high hash rate for more than a week.

\section{Blockchain Economics}\label{secEco}
In this section we will discuss economical aspects of JaxNet.

\subsection{Setting rewards in Jaxnet} \label{JaxnetRew}
In this subsection we will set BC and SC block rewards in Jaxnet. Details of this approach is discussed in the \cref{SetReward}. First, let's define coefficient $k$ from the \cref{eq:VanillaRew}:
\begin{equation}
    k= 2^{-48}\frac{\text{Jax coin}} {\text{hash}}
\end{equation}
Therefore, on BC block reward is

\begin{equation}
    R(\text{block})= D \cdot 2^{-48}\frac{\text{BC Jax coin}} {\text{ hash}}
\end{equation}
where $D=\frac{2^{256}}{T}$ is the block difficulty.

On SC block reward is

\begin{equation}
    R(\text{block})= \frac{1}{n} \cdot D \cdot 2^{-48}\frac{\text{SC Jax coin}} {\text{ hash}}
\end{equation}
where $n$ is block merged mining number discussed in the  \cref{MMproof}.

\begin{remark}
Jax coins on BC and SC may have different price.
\end{remark}

\begin{remark}
In order to curb down money creation, a slightly different formula is proposed in the \cref{InfMan}.
\end{remark}

\begin{remark}
It makes sense to set an alternative reward scheme on the Beacon chain. In this scheme the miner get an additional reward for mining BC blocks. However this reward is paid in SC coins on the next shard. The amount of this subside is calculated based on the block difficulty. This approach has multiple advantages. First, new shards has liquidity from the very beginning. The second advantage is that second advantage is incentive for miners to mine BC even if price of the coins on BC will deteriorate. Finally, it allows to set an alternative reward scheme on BC to provide scarcity for BC coins. The fixed block reward scheme used in Bitcoin could a good choice. It could give a reasonable benefit to early adopters. However, this approach has some drawback. Miners get an incentive to open new shards more often. However, to prevent this behaviour it's possible to add some extra restrictions in the network expansion protocol. 
\end{remark}

\subsection{The economics of JaxCoin}
Let us first explain the hybrid form of JaxCoin. It is not a peg, a commodity-based coin, or a fixed-supply coin that would artificially increase its market value. Our coin supply follows, nonetheless, some simple economic incentives. 

As already showed by \cite{friedman1951commodity}, the value of commodity monies are subjected to technological changes. For instance, a money is backed by oil; when new technologies such as fracking were discovered, substantial amounts of oil could be retrieved, and the price of the commodity dropped on the London and New-York markets. The same applies to JaxCoin; technological improvements of ASIC and GPU imply higher productivity of the new mining rigs. Efficiency gains in the hardware market allows miners to mine more outputs (i.e. coins) for the same amount of inputs (i.e. electricity). Everything else being equal, miners would race to have the latest mining equipment and beat the competition. Over time, this R\&D arms race \cite{alsabah2019pitfalls} will affect the price downward as miners will issue too many coins due to productivity gains in the hardware industry. We introduce below a parameter that corrects for this market externality. 

Friedman \cite{friedman1951commodity} also claims that in a commodity system, the money base would be too narrow to act as a counter-cyclical force in the economy and would just worsen price movements. However, JaxCoin does not rely on such forces. 

Linking coin supply to a ‘cost-based’ incentive mechanism has some advantages. Recall that our coin supply is proportional to the aggregated mining work executed in the network. First, we avoid backing up our coin with some fiat currencies or other baskets of assets denominated in fiat, like stable coins currently on the market. As such, Jaxcoin prices should not be as much correlated as other assets with fiat currencies. In the crypto sphere, stable coins have been a hot topic for some years as investors need some stability in prices. A lot of coins provide some mechanisms to smooth out peaks and troughs, but they are always pegged to some assets or fiat currencies that defeat the purpose of a decentralized payment system backed by mathematical rules and economic incentives in the first place. 

Since our coin is not redeemable into other assets, there is no need for JaxNet to hold on cash or collateral, which decreases the financial risks for the whole network and eases its adaptation to market needs \cite{wallace2007float}. The price of the coin will float on market exchanges according to simple supply and demand economics. Our main assumption is that miners, just like on Bitcoin, cannot mint coins at a loss. But unlike Bitcoin, supply growth is volume dependent (how much hashpower over all shards) and not time dependent (a fix reward per block). With this setting, we can avoid technical flaws and some economic flaws of cryptocurrencies, as we will see below in more details.

\subsection{Overview of the blockchain economics}
%In this section we will discuss economical aspects of the blockchain network and the reward schemes used in them.
The goal of this subsection is to prove the validity of the following proposition.
\begin{proposition}
\begin{enumerate}[a)]
\item
There is a strong reason to assume that the price of the coin is the same in each shard.
\item
Miners have an incentive to mine many shards instead of focusing on one shard because of transaction fees.
\end{enumerate}
\end{proposition}

First, we start from the observation of the simple fact: miners mine if and only if it is profitable. The incentive for miners is to mine up to the point where their expected profit equals their expected revenues minus their costs. Above this threshold, miners will not put more resources. 

Then, our coin departs also from regular institutions which hold seigniorage on money printing. Using this analogy with cryptocurrencies is slightly misleading since miners, contrary to monetary authorities, are profit oriented. This is true for existing cryptocurrencies and holds for JaxCoin. They create and make money by validating blocks. 

However, unlike conventional PoW blockchains, JaxCoin supply does not grow at a pre-specified growth rate. Instead, miners have to adjust their contribution of computing resources to both their cost structure and the new transaction needs of the network. They also have an opportunity cost in mining JaxCoin, as these resources are not put to mine other coins on other blockchains.

In the more mature coin markets using PoW algorithms, we can observe a path toward an equilibrium where mining altcoin $A$ brings a marginal profit similar to mining altcoin $B$. In disequilibrium, miners arbitrage between which altcoin is more profitable to mint. Then, in the long run, marginal profits on every altcoins mining converge back towards the equilibrium. In participating in the mining process of JaxCoin, miners have the opportunity to mine any shard. In this context, they would make an arbitrage to mine the most profitable shard. That being said, we assume that miners on shards will behave exactly like in the altcoin market: in the long run, the marginal profit will tend to be the same across all shards.

In JaxNet there are multiple shard chains which work in parallel. Miners who don't have enough resources to mine everywhere can choose on what shard chain to mine. As a result we expect that the profitability of mining on every shard chain in JaxNet will be the same. If we assume that mining expenses on some two shard chains are the same we can conclude that the expected revenues are the same. If we neglect the difference in the value of expected transaction fees we can conclude that the value of expected block rewards is the same on every shard chain. The expected number of coins in reward is the same by design.(\cref{SetReward}) Therefore, we expect that every shard coin will be valued similarly no matter on what shard it was mined. The shard coin value is about to converge toward the same value on every shard. In short, if the value of the coin on one shard will be higher than on another then we expect miners to mine on the first shard: on average mining each coin requires the same effort but the first one is more valuable. Soon values of both coins will become equal.

Let us take a closer look on the mining revenues and expenses in PoW based blockchain networks maintained in November 2019.(\cref{table:BTCcost}) Revenue often consists of two parts: transaction fees and block reward of issued coins. The main part of the expenses is purchasing of the mining hardware, bills for the electricity, network connection and data storage. This is true for the majority of PoW based blockchain networks as well as JaxNet.

The \cref{BTCbreakdown} contains an estimates by Croman et al. on distribution of miners expenses over 5 main categories.\cite{croman2016scaling} The general pattern is that mining hardware and electricity bills constitute around 98\% of overall expenses. These expenses are strikingly high and generally considered as a waste of resources since common payment systems do not have them.  
\begin{table}[ht] \label{table:BTCcost}
\centering
\captionsetup{width=\columnwidth,font=footnotesize,justification=centering}
  \caption{Bitcoin cost breakdown. Includes cost incurred by all nodes.\cite{croman2016scaling}}
  \label{BTCbreakdown}
{
\def\arraystretch{1.5}%
\resizebox{0.95\columnwidth}{!}{
\begin{tabularx}{1.05\columnwidth}{@{\extracolsep{\fill}}Xllll}
\toprule
%\hline
 & \multicolumn{2}{c}{at max troughput} & \multicolumn{2}{c}{at \textit{de facto} troughput}\\ \cline{2-3} \cline{4-5}
 & cost/tx & percentage & cost/tx & percentage \\
\midrule
Mining: proof-of-work   & $\sim \$0.8 - \$1.7$ & $\sim 56\%$ & $\sim \$3.6$   & $\sim 56\%$ \\
Mining: hardware        & $\sim \$0.6 - \$1.3$ & $\sim 42\%$ & $\sim \$2.7$   & $\sim 42\%$ \\
Transaction validation  & $\sim \$0.002$       & $\sim 0.2\%$& $\sim \$0.008$ & $\sim 0.2\%$\\
Bandwidth               & $\sim \$0.02$        & $\sim 2\%$  & $\sim \$0.08$  & $\sim 2\%$  \\
\hline
Storage (running cost)  & \multicolumn{4}{c}{$\sim \$0.0008/5$ years} \\
\bottomrule
\end{tabularx}}
}
\end{table}

Let's take a look on the distribution of miners' revenues. In networks like Bitcoin around 99\% of reward comes from coin issuance and only 1\% from transaction fees. Transaction fees are determined by the demand on the block body space and limited proposal of such space. Users are interested to keep their transaction fees as minimum as possible. We can expect that in scalable blockchain network with high throughput transaction fees will be minimal.\cite{lavi2019redesigning} However, in Jaxnet transaction fees have an important role and their value should be noticeable. One possible approach is to set a lower bound on the transaction fee. However, the paper by Lavi et.al.\cite{lavi2019redesigning} describes a promising fix to this problem.

In JaxNet miners have the opportunity to choose on what shard to mine. Let's make an estimation of the possible profit or lose from their decision to mine the shard $i$. The mathematical expectation of the block reward of issued coins doesn't depend on this choice. On the other hand miners could increase the expected value of the second part of their reward. Expected reward of miner from transaction fees is given by the formula:
\begin{equation}
    E(\text{reward})=\sum_{i}TF_i \cdot \frac{MHR}{SHR_i+MHR}
\end{equation}
where $TF_i$ is aggregate transaction fees of transactions included into the shard block, $MHR$ is particular miners hash rate, $SHR_i$ is the shard $i$-th hash rate excluding particular miner.

Let $\epsilon_i$ to be the cost of processing $i$-th shard. Then 
\begin{equation}
\begin{split}
    \triangle_i=\frac{TF_i\cdot MHR}{SHR_i+MHR}-\epsilon_i= \\
    =\frac{TF_i}{1+\frac{SHR_i}{MHR}}-\epsilon_i
\end{split}
\end{equation}
is the miners possible profit if he decides to join mining on $i$-th shard. One can observe that the higher his hash rate the higher value of this expression. Therefore, miners or mining pools with higher hash rate has more incentive to mine more shards. Also miners who mine many shards have an incentive to increase their revenue by increasing their hash rate. Another catch is that shards higher average sum of transactions fees $TF_i$ are more likely to attract miners. So we expect that shards with more intensive trading will be better protected.
\subsection{Excessive coin supply?}
In this subsection we are going to discuss money creation and its potential impact for the value of JaxCoins. First, we need to give some definitions that will apply throughout the rest of the paper. 

First, we need to distinguish \textit{monetary creation} from \textit{monetary mass}.The later is the total of coins minted since the genesis block. The former is the increase in money supply within a time interval. In the blockchain network the measure of time is the number of blocks in the chain. The measure of money are coins. \textit{Coin supply} at the block number $n$ is the number of coins issued from the beginning of the chain up to the block number $n$ inclusively. In the setting of JaxNet we call monetary creation from block $n$ to block $m$ be the fraction:
\begin{equation}
    M_S(n,m)=\frac{\text{Coin supply at the block \#m}}{\text{Coin supply at the block \#n}}-1
\end{equation}
where $n<m$ and $M_S$ is the notation for the monetary creation through this paper.

To keep the things simple we do not distinguish the coins in active circulation from coins on so called "cold accounts". 
%In the case of Bitcoin there is a study that 30\% of coins may have been lost forever. 
Monetary creation defined above has an advantage that it could be easily calculated from the blockchain history. JaxNet consists of multiple shards and multiple chains. Therefore, the Beacon chain and each shard chain has its own monetary mass and monetary creation. However, by the design of the reward scheme we can expect that the coin supply on the Beacon chain is nearly equal to the aggregate coin supply of shards. This fact makes monetary creation on the Beacon chain a good measure of the newly minted coins in JaxNet.

In JaxNet monetary creation is directly related to the security against 51\% attacks (see \cref{SecMod}).

There is no sense in coins which do not have an exchange value. Let's assume that Beacon chain coins and shard coins are used as exchange tokens in some market. Assume on this market there is a supply and demand, coins have a constant velocity of circulation and some market price or some equivalent of goods at the target moment. The price of the coin may vary at different moments of the time. The price inflation is the tool for measuring this difference.  

Bitcoin and the majority of other blockchain networks have a limited supply of coins. This involves scarcity of the coins in the market. Thus the price on them could fluctuate up and down. However, in such systems coin creation decreases to zero as time passes. This sort of monetary policy can cause problems in the future. \cite{BTCsecbud1,BTCsecbud2,BTCdisToDeath} In JaxNet the coin supply is not limited. However, the proposed design involves positive monetary creation which is discussed in the \cref{PrInfl}. The possible impact of money creation on blockchain economics is discussed in \cref{ShortTermRisks}.

One may think that giving the reward proportional to the effort is a bad idea since miners will mint huge amount of coins so that further mining will be unprofitable. We can argue that there are three forces which could drive the further mining. First of them is technological progress. Launches of more efficient new mining rigs on the market could intensify mining in the network. Whenever such event happens, mining in the network is boosted during a certain interval of time. Let's make a few estimates. Assume that the situation on the market is stable so that demand and supply remain at the same level and velocity of the currency is constant. Assume that minting the coin on the new device is $\gamma$ times more cost efficient than on the previous efficiency leader. Let the previous volume of currency on the market is $V$ coins and further minting of coins was nearly unprofitable. Then we can expect that the new equilibrium on this market will be reached once coin supply will reach $\gamma \cdot V$. Therefore
\begin{equation}
(\gamma-1)V
\end{equation}
coins are going to be minted. One may suppose that this will entail a dramatic emission of coins within a short period. There is indeed an upper bound in the production of mining equipment that cannot satisfy all the demand in the short term. Besides, we argued that miners' investment is irreversible, so that they can delay their investment decisions. They are more likely to do so if they anticipate just short-lived productivity gains due to new equipment and not sustainable demand. Another economic dimension to take into account is that PoW mining is an opportunity cost, meaning that miners have to give up on resources that could be used for other profitable endeavours, being mining other PoW blockchains or something else altogether. This opportunity cost is correlated to the monetary supply and its exchange rate \cite{choi2019money}, meaning that miners will switch their activity if their expected payoffs go down due to oversupply or price crashes. This will reduce the hash rate across all shards and therefore the coin supply. Finally, recall that miners maximize their profit in aligning with the quantities produced by other mining pools. Marginal profit will soon reach zero and supply will stabilize, everything else being equal.

The second force which supports our coin issuance mechanism is the fact that tokens sometimes get lost. It's caused by hardware issues, inaccuracy, death of their owners and token burning schemes\cite{BurnCoin}. For example, there is an estimate that nearly $30\%$ of issued BTC are already lost.\cite{30BTClost} Lost tokens no longer participate in coin circulation and get replaced by newly issued coins.

The third force is the growth of the market. Whenever this happens there is always need for new coins for transaction purposes.

Technological improvements in mining hardware inevitably cause price decrease. So it's important to have an insight on the rate of this process. An attempt to get an estimate is given in the \cref{PrInfl}. Coin devaluation is not something token holders want to see. It is good to have a mechanism which takes into account these efficiency gains to correct downward movements of prices. Our proposal is discussed in the \cref{InfMan}.
\subsection{Coin creation management} \label{InfMan}
Our coin supply management is inspired by rule-based monetary economics. Although these policies are not implemented in today’s central banks policies, they are relevant to manage the coin supply efficiently in Jax Network.

Furthermore, JaxCoin is not affected by the type long-term deflationary problems that are inherent to Bitcoin or other commodity monies, because coin supply will adapt to demand through a non-cooperative game as we will see below. Ultimately, the fundamental value of our coin is correlated to the costs of mining, much closer than Bitcoin can be. Besides, the implementation of a constant further contracts the money supply. The main objective of the protocol is to provide a stable coin with simple issuance rules inspired by the monetary economics of Friedman \cite{friedman1960program} and Taylor \cite{taylor1993discretion}. Both authors, in different ways, suggested a constant rate of money supply irrespective of business cycles. 

Below in this section we will discuss possible tools for coin supply management within JaxNet. As we have learned from the \cref{PrInfl} money creation can affect prices in the nearest future, and therefore destroy the fundamental value of the coin. Actually, economists often state that currency serves as a medium of exchange and as a store of value. Following the quantitative theory of money, high level of coin issuance would undermine the currency as a store of value, everything else being equal.

Economic setting and assumptions are discussed in the \cref{Economic assumptions}.

To avoid currency crashes and high volatility (\cref{ShortTermRisks}), coin issuance should obey a strict set of rules so that miners can anticipate their expected payoffs and adjust their hashrate accordingly.

\subsubsection{Naive approach and obstacles}
One can set mining epochs so that block reward will decrease with time in some way. As always, the measure of the logical time in the system is the length of the chain. Each shard has its length which determines its mining epoch. If one properly set mining epochs and rewards within them, he will be able to address high coin supply. However, there are a few obstacles on this way.

First, monetary policy for coin issuance should follow a clear and transparent set of rules. If monetary creation is high it could lead to high price inflation. If it is too low it could provoke emergence of an economic bubble on the blockchain and dangerous fluctuations of the hash rate. Desired mechanism should properly work over the long term, and be simple enough for miners to calculate their expected profits. It is possible that in the future technological advancement will slow down and won't bring much yearly improvement in energy efficiency of semiconductors. Monetary creation shouldn't drop to zero and cause security issues anticipated for bitcoin-style monetary policy.  

Second, some synchronization mechanism is required to make sure that shard chains follow the same rules. Timestamps, block reference mechanisms could be falsified by malicious actors. The length of the chain is a good substitute for the time within a single chain. However, here we face multiple problems. First, chain length is not a precise measure of time. Basically, chain length is random variable with some distribution. We can only shape the parameters of this distribution. Second, casual miners and malicious actors can try to manipulate timestamps or accept wrong timestamps if they find it beneficial. Moreover, in the setting of JaxNet there are multiple chains and every chain has its own length.

\subsubsection{Proposal of the  mechanism}
Obstacles mentioned above make the task rather tricky. Nevertheless, we offer a solution below that fixes part of the aforementioned issues. 

First, let's mention some useful features which JaxNet already has.
\begin{enumerate}[a)]
    \item
    One can expect that an overwhelming majority of miners mine BC. It doesn't cause any extra workload and brings profit. If all miners in JaxNet mine on BC then we can expect that the coin supply on BC approximately equals to aggregate the coin supply on shard chains. Thus, the study of BC provides us an insight on intensity of mining and coin issuance in the whole network.
    \item Calculation of aggregate block difficulties on BC within a certain interval is easy and provides an estimate of average hash rate within that interval.
    \item Difficulty and timestamp control rules in the \cref{DiffandTimeS} are designed to keep chain length and real time almost synchronous.
\end{enumerate}

Let's split every blockchain into sequences of mining epochs. Each mining epoch on the chain is a sequence of blocks. Limits of every epoch are determined by chain length.

On the BC first mining epoch starts with the genesis block \#$0$. Mining epoch \#$m$ ends with the block \#$(L \cdot m)$ and then next epoch starts. We set $L$ to be $4096$.

On every SC, the last block in the epoch is determined similarly. The genesis block of SC is some block on BC and its epoch is already defined. The difference is that every block on BC corresponds to $40$ blocks on $SC$. So in order to "close" an epoch forty times more SC blocks are required.

We propose to set a reward for the block $B_n$ in the mining epoch \#$m$ according to the next formula:
\begin{equation}
R(B_n) = D(B_n) \cdot \prod_{i=1}^{m-1}k_i  
\end{equation}
Here $R$ and $D$ are reward and difficulty respectively. $k_i$ is an \textit{adjustment coefficient} of the mining epoch $i$. Adjustment coefficient $k_i$ is defined as follows. We set
\begin{equation}
\lambda=\exp({\ln(0.8)/144})=0.99845159
\end{equation} 
Then for the first mining epoch we set
\begin{equation}
k_1=\lambda^{12}
\end{equation}
Then we follow next rules. We calculate aggregate difficulty in $i$-th epoch $D_i$ on BC and aggregate difficulty of BC $D$ from first epoch to $i$-th epoch. 
\begin{equation}
D=\sum_{j=1}^{i}D_j  
\end{equation}
\begin{equation}
    k_i=
    \begin{cases}
        k_{i-1} & \text{if } \frac{D-D_i}{D}<\lambda^{4}\cdot k_{i-1}^{3/2}\\
        k_{i-1}\cdot \lambda^{-2} & \text{if }
        \left(\frac{D-D_i}{D} \ge \lambda^{2}\cdot k_{i-1}^{3/2}\right) \wedge (k_{i-1}<1) \\
        1 & \text{if } k_{i-1}=1
    \end{cases}
\end{equation}
Motivation of the rules above is the following. They somewhat reduce coin issuance and make coins more scarce. If there is insufficient mining in the network or a sign that mining will be reduced, then adjustment coefficients get larger. These should stimulate mining in next epochs by making it more profitable.

Notice that because of the difficulty control rules described in the \cref{DiffandTimeS} one can calculate adjustment coefficient of the epoch before its end. On average there should be two weeks in reserve when next mining epoch in the shard starts. Therefore, one can almost always calculate how many coins were issued in the certain block even if it was minted recently.

\subsubsection{Some economic considerations}

Overall, unlike conventional PoW blockchains, JaxCoin supply does not grow at a pre-specified growth rate. Instead, miners must adjust their supply to both their cost structure and the new transaction needs of the network. 

Our coin supply management is inspired by rule-based monetary economics to constrain our monetary supply. Although these policies are not implemented in today’s central banks policies, they are relevant to manage the coin supply efficiently in Jaxnet. Besides, we do not apply these rules right on. As we have seen they are amended to fit the technical aspects of blockchain-based protocols. 

Scarcity of the money supply is economically bounded by the fact that a miner will not allocate resources if it is not profitable for him. Money supply, although theoretically unlimited, cannot grow above the cost of energy and the productivity gains observed in the hardware mining equipment. The theoretical growth rate is parametrized by a constant $k$. This is an upper bound to which every miner needs to abide by, which is correcting for efficiency gains in mining equipment. A daily feed will be calculated based on economic forces and provided to every mining pool. This practical $k$ cannot be higher than the theoretical $k$. Miners will be able to choose to apply any $k$.\footnote{We assume here that miners are in a Cournot game played indefinitely; they choose quantity according to other miners. Therefore, the more mining pools, the higher the welfare. }

\section{Conclusions}\label{secCon}
\subsection{Addressing Scalability Trilemma}
In this subsection we are going to summarize how our solution addresses Blockchain Scalablity Trilemma. The main question is whether this problem is solved or how well it is addressed. Since there is no unique description of Trilemma and the community may have different expectations there is a place for speculations on this subject.

First, let us start from the scalability part of the Trilemma. One may argue that any sharding scheme is a stealth increase of the block size, and our solution is kind of that. Really, if a given node decides to become the fullest node in the network and operate all shards that will be the case. The storage and bandwidth requirements for him will grow almost linearly with the number of shards. Actually, the scheme proposed in this paper requires a number of such nodes or mining pool operators. However, not all nodes in the network are obligated to do that. Our solution provides a freedom of choice. Nodes can become full nodes for the specific shard of few specific shards without significant penalties to their margins. Moreover, in any solution with a decentralized nature, transactions should be processed by multiple nodes. Our solution assumes that some nodes do the most part of that work. The question is, how many nodes are able to achieve that level of performance? Data storage requirements and transaction validation according to \cref{BTCbreakdown} is the cheapest part. The bandwidth requirement could be the problem. However, even today 40GBit internet cables are not rare. Setting a good reliable connection between most heavy nodes does not seem an unsolvable problem. It is affordable compared to expenses on mining hardware.

%Another argument against the scalability is the fact that cross-shard transactions rely on the third party. The author of this paper admits that the best solution to this issue is yet to be found. Cross-shard and cross-chain interaction is a well-known area of active research.

Second, let us discuss the decentralization part of the Trilemma. One may argue that in our solution there are a limited number of powerful nodes which together have overwhelming influence on the blockchain network. The authors of this paper admit that coalitions of powerful nodes pose a threat of centralization. However, if we take a close look at existing decentralized solutions like Bitcoin and Ethereum we observe that a major part of the computational power is aggregated by a few mining pools.\cite{samani2018models} Participants of those pools are not involved neither in block creation or block verification. They simply perform blind mining. Nevertheless, participants are free to join and leave the mining pool if it does something inappropriate or is about to aggregate hash rate power up to a critical point. Although, as a rule, mining pool operators are interested in maintaining the network and conducting mining in accordance to the protocol requirements. We can conclude the dominating role of mining pools is considered to be acceptable.

Third, security side of the Trilemma is not discussed yet. This part is discussed 
%in details 
in the \cref{SecMod}. Commonly, the hash rate is considered to be the measure of the security against $51\%$ attacks. It works fine when we compare networks with identical design. However, researchers in order to achieve scalability often propose some Proof-of-Stake consensus schemes. These designs do not involve mining as generating a hash. Thus, it is not obvious how to make a direct comparison. Nevertheless, security always ascend to the money. In order to perform an attack, a malicious actor needs to make an investment of money of the same magnitude as one that network pays for its security. We argue that JaxNet provides a good trade-off between the security and other parameters of the system.

One parameter in the \cref{ExpNet} is of particular interest. We create a new shard when roughly $75\%$ of the network hash rate votes for that. As the system grows we make a trade-off between scalability, security and decentralization. If we expand too aggressively the network becomes less secure and less decentralized. On the other hand, establishing new shards improves scalability.

\subsection{Open questions for further research}
Here we present the list of most interesting open questions related to JaxNet which will be interesting for the further research.

\begin{question}
Are there better solutions to the blockchain Scalability Trilemma?
\end{question}

\begin{question}
Is there a reliable and efficient way to set cross-shard transactions in JaxNet protocol?  
\end{question}
Cross shard transaction solutions are an area of extensive research. Although many approaches were proposed in the past few years a good solution for issue is yet to be found.

\begin{question}
Is there a better way to organize consensus between nodes on the total number of shards?
\end{question}

It's a tricky question. On the one hand the network needs scalability in order to effectively process more transactions. Casual users are interested in numerous shards with reliable and cost efficient transactions. On the other hand miners are not much interested in increasing the number of shards since they can charge high fees on the limited number of shards. Also increasing the number of shards to the level where only few miners will be able to support majority of them is big threat to the security and decentralization.

\begin{question}
Is there a wise way to control money creation in decentralized network? 
\end{question}

On the one hand it's good for coin holders when the coin creation rate is low. On the other hand issuance of new coins in JaxNet is the payment for the security against $51\%$ attacks in shards. It's possible to set more complex time dependent reward functions which reduce the reward as a number of blocks on the main chain growth similar to Bitcoin. In this case mining in JaxNet will be less profitable. If at some point in the future technological advancement due to Moore's law will become slower there is a possibility that honest mining will be less profitable and hash-rate in the system will dramatically fail. In this situation malicious actions could become more profitable and poses a serious thread to the system.

\begin{question}
Is there a way to reorganize existing blockchain networks and improve their scalability in the fashion similar to JaxNet?
\end{question}

The authors of this paper believe this is possible. However major improvements to their protocol should be adopted. Moreover, after such hard forks they may lose some of their key features. Also the baggage of the previous transaction history could make them less efficient than JaxNet.

%%%\bibliographystyle{unsrt}
%%%\bibliographystyle{none}

%\bibliographystyle{plain}
%\bibliography{blockchain}
%\newpage
\printbibliography

\newpage
\appendix

\section{Architecture details}
\subsection{Merkle-proofs} \label{AppMT}

Merkle trees and Merkle proofs are backbones of the blockchain technology. They were patented by Ralph Merkle in 1979 and became widely used along with similar digital signature schemes. \cite{merkle1987digital} Merkle Tree has a binary tree structure with data chunks at the bottom below tree leaves (\cref{Merkle1}). Merkle tree of height $h$ has $2^h$ leaves. In order to prove integrity of certain data on constructs Merkle Proof (\cref{Merkle2}) of length $h$ hashes (not including Root). Merkle Proof could bee verified after $2h$ hash calculations.
\begin{figure}[!ht]
\centering
%\begin{minipage}{.5\textwidth}
\begin{subfigure}{.48\textwidth}%[ht!]
	\begin{center}
		\resizebox{\linewidth}{!}{
			\begin{tikzpicture}[
			%node distance=20mm,
			level distance=80pt,
			level 1/.style = {sibling distance=220pt},
			level 2/.style = {sibling distance=110pt},
			level 3/.style = {sibling distance=55pt},
			every node/.style=MerkleNode,
			every child/.style={mychild}]
\node(a){Root}
    child foreach \x in {0,1}
        {node(H\x){$H_{\x}$}
        child foreach \y in {0,1}
            {node{$H_{\x\y}$}
            child foreach \z in {0,1}
                {node{$H_{\x\y\z}$}
                child {node[MerkleData]{DATA}}
    }}};
			\end{tikzpicture}
		}
		\vspace*{3pt}
		%\captionsetup{labelformat=empty}
		\caption{\textbf{Merkle Tree}} \label{Merkle1}
    \end{center}
\end{subfigure}
%\end{minipage}
%\begin{minipage}{.5\textwidth}
\begin{subfigure}{.48\textwidth}%[ht!]
	\begin{center}
		\resizebox{\linewidth}{!}{
			\begin{tikzpicture}[
			%node distance=20mm,
			level distance=80pt,
			level 1/.style = {sibling distance=220pt},
			level 2/.style = {sibling distance=110pt},
			level 3/.style = {sibling distance=55pt},
			every node/.style=MerkleNode,
			every child/.style={mychild}
			]
\node[MerkleRoot] {Root}
    child[draw=blue!80] {node {$H_0$}
        child[draw=green!80] {node[MerkleProofNode] {$H_{00}$}
            child {node {$H_{000}$}}
            child {node {$H_{001}$}}
        }
        child[draw=blue!80] {node {$H_{01}$}
            child[draw=blue!80] {node {$H_{010}$}
                child[draw=blue!80] {node[MerkleProofData]{DATA}}
            }
            child[draw=green!80] {node[MerkleProofNode] {$H_{011}$}}
        }
    }
    child[draw=green!80] {node[MerkleProofNode] {$H_1$}
        child {node {$H_{10}$}
            child {node {$H_{100}$}}
            child {node {$H_{101}$}}
        }
        child {node {$H_{11}$}
            child {node {$H_{110}$}}
            child {node {$H_{111}$}}
        }
    };
			
%\node%[MerkleRoot]
%(b) 
%{Root}
%    child {\node%(bH0)
%    {$H_{0}$}}
%    child {\node%(bH1)
%    {$H_{1}$}
        %child 
            %{node(bH00){$H_{00}$}
            %child 
                %{node(bH000){$H_{000}$}
                %child {node[MerkleData]{DATA}}
%    };
			\end{tikzpicture}
		}
		\vspace*{3pt}
		%\captionsetup{labelformat=empty}
		\caption{\textbf{Merkle Proof}} \label{Merkle2}
    \end{center}
\end{subfigure}
%\end{minipage}
        \vspace*{3pt}
		%\captionsetup{labelformat=empty}
		\caption{\textbf{Merkle}} \label{Merkle}
\end{figure}
In JaxNet Merkle trees are used to efficiently keep integrity of transactions in the body block. Another instance of usage is Shard Merkle tree. In each instance SHA-3 hashing algorithm is used in order to achieve better performance. Therefore the size of the Merkle proof in the solution is often $8h$ bytes. 
\subsection{Tree Encoding}\label{TreeEnc}
The number of full binary trees with $n+1$ leaves is $n$-th Catalan number $C_n$. Catalan numbers and various counting problems in which they appear are well studied in  Combinatorics.\cite{lando2003lectures, roman2015introduction,davoodi2017succinct} There is a one-to-one map between full binary trees with $n+1$ leaves and Dyck words of length $2n$. Also it's known as correspondence between full binary trees and balanced parenthesis. We use this correspondence to encode orange subtree as a sequences of bits. Therefore orange subtree with at most $n+1$ leaves will be encoded with $2n$ bits. However, such encoding always starts from '1' or from open parenthesis '('. It also ends with "0" which represents ")". We drop last symbol and get encoding with $2n-1$ bits. Therefore, merge mining tree encoding (\cref{tikz:TreeEnc}) always have $3n$ bits. 

\def\TreeTop{1,1,0,0,1,1,0,0,1,0,0}
\def\HashTypes{0,1,0,1,1,0}
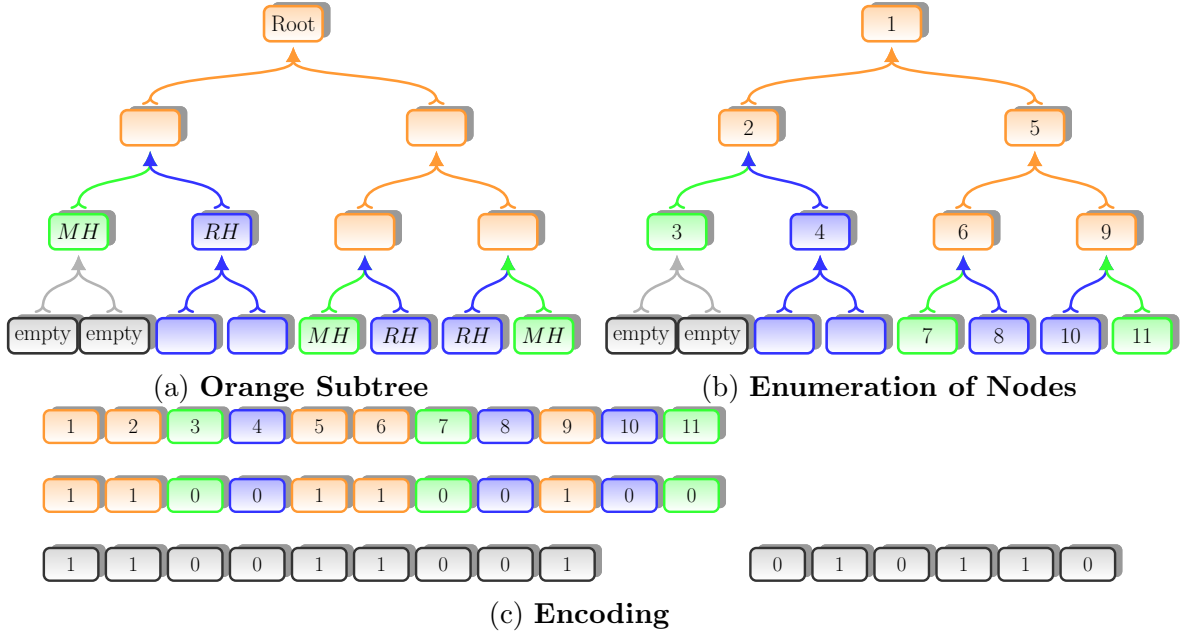
\begin{figure}[!ht]
\centering
\begin{subfigure}{.48\textwidth}%[ht!]
	\begin{center}
		\resizebox{\linewidth}{!}{
			\begin{tikzpicture}[
			level distance=80pt,
			level 1/.style = {sibling distance=220pt},
			level 2/.style = {sibling distance=110pt},
			level 3/.style = {sibling distance=55pt},
			every node/.style=MerkleNode,
			every child/.style={mychild}]
\node[MerkleOrangeNode] {Root}
    child[draw=orange!80] {node[MerkleOrangeNode] {}
        child[draw=green!80] {node[MerkleProofNode] {$MH$}
            child[draw=gray!60] {node {empty}}
            child[draw=gray!60] {node {empty}}
        }
        child[draw=blue!80] {node[MerkleBlueNode] {$RH$}
            child[draw=blue!80] {node[MerkleBlueNode] {}
                %child[draw=blue!80] {node[MerkleProofData,
                %rectangle split parts=2]{Shard \\ block}}
            }
            child[draw=blue!80] {node[MerkleBlueNode] {}
                %child[draw=blue!80] {node[MerkleProofData,
                %rectangle split parts=2]{Shard \\ block}}
            }
        }
    }
    child[draw=orange!80] {node[MerkleOrangeNode] {}
        child[draw=orange!80] {node[MerkleOrangeNode] {}
            child[draw=green!80] {node[MerkleGreenNode] {$MH$}}
            child[draw=blue!80] {node[MerkleBlueNode] {$RH$}
                %child[draw=blue!80] {node[MerkleProofData,
                %rectangle split parts=2]{Shard \\ block}}
            }
        }
        child[draw=orange!80] {node[MerkleOrangeNode] {}
            child[draw=blue!80] {node[MerkleBlueNode] {$RH$}
                %child[draw=blue!80] {node[MerkleProofData,
                %rectangle split parts=2]{Shard \\ block}}
            }
            child[draw=green!80] {node[MerkleGreenNode] {$MH$}}
        }
    };
			
            \end{tikzpicture}
		}
		%\vspace*{3pt}
		\caption{\textbf{Orange Subtree}} \label{Merkle4}
    \end{center}
\end{subfigure}
\begin{subfigure}{.48\textwidth}%[ht!]
	\begin{center}
		\resizebox{\linewidth}{!}{
			\begin{tikzpicture}[
			level distance=80pt,
			level 1/.style = {sibling distance=220pt},
			level 2/.style = {sibling distance=110pt},
			level 3/.style = {sibling distance=55pt},
			every node/.style=MerkleNode,
			every child/.style={mychild}]
\node[MerkleOrangeNode] {1}
    child[draw=orange!80] {node[MerkleOrangeNode] {2}
        child[draw=green!80] {node[MerkleProofNode] {3}
            child[draw=gray!60] {node {empty}}
            child[draw=gray!60] {node {empty}}
        }
        child[draw=blue!80] {node[MerkleBlueNode] {4}
            child[draw=blue!80] {node[MerkleBlueNode] {}
                %child[draw=blue!80] {node[MerkleProofData,
                %rectangle split parts=2]{Shard \\ block}}
            }
            child[draw=blue!80] {node[MerkleBlueNode] {}
                %child[draw=blue!80] {node[MerkleProofData,
                %rectangle split parts=2]{Shard \\ block}}
            }
        }
    }
    child[draw=orange!80] {node[MerkleOrangeNode] {5}
        child[draw=orange!80] {node[MerkleOrangeNode] {6}
            child[draw=green!80] {node[MerkleGreenNode] {7}}
            child[draw=blue!80] {node[MerkleBlueNode] {8}
                %child[draw=blue!80] {node[MerkleProofData,
                %rectangle split parts=2]{Shard \\ block}}
            }
        }
        child[draw=orange!80] {node[MerkleOrangeNode] {9}
            child[draw=blue!80] {node[MerkleBlueNode] {10}
                %child[draw=blue!80] {node[MerkleProofData,
                %rectangle split parts=2]{Shard \\ block}}
            }
            child[draw=green!80] {node[MerkleGreenNode] {11}}
        }
    };
			
            \end{tikzpicture}
		}
		\caption{\textbf{Enumeration of Nodes}} \label{Merkle5}
    \end{center}
\end{subfigure}
\begin{subfigure}{.9\textwidth}%[ht!]
	\begin{center}
		\resizebox{\linewidth}{!}{
			\begin{tikzpicture}[
			level distance=80pt,
			level 1/.style = {sibling distance=220pt},
			level 2/.style = {sibling distance=110pt},
			level 3/.style = {sibling distance=55pt},
			every node/.style=MerkleNode,
			every child/.style={mychild}]
%\foreach \x in {1,2,3,4,5,6,7,8,9,10,11}			
%{\node[MerkleGreenNode] (BC\x)
%at ({\x*1.8},0) {\x};}
%\foreach \x[count=\xx from 2] in {1,2,3,4,5,6,7,8,9,10}
%{\draw[myarrow] (BC\xx) -- (BC\x);}

\foreach \xx[count=\x from 1] in \TreeTop %{1,2,3,4,5,6,7,8,9,10,11}
{
\ifthenelse{\xx=1}{
  \node[MerkleOrangeNode] (U\x)
  at ({\x*1.8},0) {\x};
  \node[MerkleOrangeNode] (D\x)
  at ({\x*1.8},-2) {\xx};
  \node[MerkleNode] (D\x)
  at ({\x*1.8},-4) {\xx};
}{
  \ifthenelse{\x=4 \OR \x=8 \OR \x=10}{
    \node[MerkleBlueNode] (U\x)
    at ({\x*1.8},0) {\x};
    \node[MerkleBlueNode] (D\x)
    at ({\x*1.8},-2) {\xx};
    \ifthenelse{\x <10}{
        \node[MerkleNode] (D\x)
        at ({\x*1.8},-4) {\xx};
    }{}
  }{
    \node[MerkleGreenNode] (U\x)
    at ({\x*1.8},0) {\x};
    \node[MerkleGreenNode] (D\x)
    at ({\x*1.8},-2) {\xx};
    \ifthenelse{\x <10}{
        \node[MerkleNode] (D\x)
        at ({\x*1.8},-4) {\xx};
        %\draw ({\x*1.8},-4) \node[cross,red] {};
    }{}
  }
}
}
\foreach \xx[count=\x from 1] in \HashTypes
{
    \node[MerkleNode] (D\x)
    at ({20.5+\x*1.8},-4) {\xx};
}
            \end{tikzpicture}
		}
		%\vspace*{3pt}
		\caption{\textbf{Encoding}} \label{Merkle6}
    \end{center}
\end{subfigure}
        \vspace*{1pt}
		\caption{\textbf{Orange subtree encoding}} \label{tikz:TreeEnc}
\end{figure}

In JaxNet orange subtrees have limited size. If say that orange with $n+1$ leaves has the size $n+1$. Size limit depends on the height $h$ of the Shard Merkle Tree and given in the \cref{tableEnc1} below.

\begin{table}[ht]
\centering
\captionsetup{width=\columnwidth,font=footnotesize,
  justification=centering}
  \caption{Size of Merge-mining tree encoding}
  \label{tableEnc1}
{
\def\arraystretch{1.5}%
\resizebox{0.75\columnwidth}{!}{
\begin{tabularx}{0.9\columnwidth}{@{\extracolsep{\fill}}>{\raggedright\arraybackslash}Xlll@{}} 
\toprule
\makecell[l]{Height of\\ the Shard \\ Merkle tree} &
\makecell[l]{Bound on the \\ max size of \\ orange subtree}&
%\makecell[l]{Number of\\ possible \\ orange subtrees} &
\makecell[l]{Max size of \\ orange subtree \\ encoding in bits}&
\makecell[l]{Max size of \\ Merge-mining tree\\ encoding in bits}\\
\midrule
1 & 2  & 3 & 3\\
2 & 4  & 7 & 9\\
3 & 8  & 15 & 21\\
4 & 16  & 31 & 45\\
5 & 30  & 59 & 90\\
$h>5$ & 6h  & 12h-1 & 18h\\
20 & 120  &239& 360\\
\bottomrule
\end{tabularx}}
}
\vspace{.2cm}
\end{table}

Block sizes increase as number of shards in the network grows. Relevant data is aggregated in the \cref{tableEnc2}.

\begin{table}[ht]
\centering
\captionsetup{width=\columnwidth,font=footnotesize,
  justification=centering}
  \caption{Size of block components}
  \label{tableEnc2}
{
\def\arraystretch{1.5}%
\resizebox{0.95\columnwidth}{!}{
\begin{tabularx}{1.15\columnwidth}{@{\extracolsep{\fill}}>{\raggedright\arraybackslash}Xlllll@{}} 
\toprule
\makecell[l]{Height of\\ the Shard \\ Merkle tree} &
\makecell[l]{Max size of \\ MM tree \\ encoding, bits}&
\makecell[l]{Max size of \\ shard \\ proof, bytes}&
\makecell[l]{Max size of \\ merge-mining \\ proof, bytes}&
\makecell[l]{Max size of \\ BC block \\ header, bytes}& 
\makecell[l]{Max size of \\ SC block \\ body, KB}\\
\midrule
1 & 3 & 32 & 32 &121& $\sim32$\\
2 & 9 & 64 & 64 &121& $\sim32$\\
3 & 21 & 96 & 128 &122& $\sim32$\\
4 & 45 & 128 & 256 &126& $\sim32$\\
5 & 90 & 160 & 960 &132& $\sim32$\\
$h>5$ & 18h & 32h & 192h& $120+\lceil 9h/4 \rceil$& $\sim32$\\
20 & 360 & 640 & 3840 &165& $\sim32$\\
\bottomrule
\end{tabularx}}
}
\vspace{.2cm}
\end{table}

In fact, finding the number of full binary trees with at $n+1$ leaves and height less than $h$ is another well-known problem in Combinatorics and Probability theory. Application of reflection principle gives the formula to calculate the number mentioned above:
\begin{equation}
    \binom{2n}{n}-\binom{2n}{n-1}+\binom{2n}{n+2h}-\binom{2n}{n+2h-1}
\end{equation}
Since we take $n=6h$ it becomes
\begin{equation}
    \binom{12h}{6h}-\binom{12h}{6h-1}+\binom{12h}{8h}-\binom{12h}{8h-1}
\end{equation}
which could be approximated as
\begin{equation}
    \simeq C_{6h} \simeq \frac{2^{12h}}{\sqrt{\pi}(6h)^{3/2}}O\left(1+\frac{1}{h}\right)
\end{equation}
Even for rather small values of $h$ it is a huge number. This estimate reflects the fact that, despite some limitations, we provide a huge flexibility for selecting the subset of shards for merged mining.

It is possible to have an empty orange subtree. In this case encoding has $0$ bits. If Shard Merkle Tree root is magic then any shard was merge-mined. If it's not a magic hash then it is assumed that all open shards were merge-mined.
\section{Discussion of the FlyClient}\label{FlyBugs}
This subsection is devoted to FlyClient introduced by B{\"u}nz \textit{et al.}\cite{bunz2019flyclient}. The original paper with its description contains few flows and the misleading argument. Since this construction plays an important role in our design it's reasonable to investigate the issue. %The main problem is that the author has chosen inappropriate consensus rules and difficulty transition rules.

\subsection{Flaw in the construction and consensus}
\subsubsection{Description of the problem}
The problem could be described as follows. 
1

The author of the original design claims that the proposed super-light client should verify all difficulty transitions in the chain. This verification requires the data about timestamps and block difficulties which is often stored in the block headers. The simplest way to get access to this data and verify its integrity is to download necessary block headers. However, the main purpose of the super-light client is to reduce the number of headers which should be downloaded. Let's find answers for the next three questions:
\begin{enumerate}
\item
How much data is required to be downloaded in the original design of the FlyClient in order to verify all difficulty transition?
\item
Does verification of difficulty transition rules in the original design of the FlyClient work as intended and reach its purpose?
\item
Is there any need in this verification?
\end{enumerate}

B{\"u}nz \textit{et al.} consider a model in which difficulty adjustment occurs every $N$ blocks. Although difficulty adjustment algorithm vary from one blockchain protocol to another, a timestamp from at least one block from each $N$-block interval is required. Therefore, the number of block headers required for verification of every difficulty transition within the chain linearly depends on the length of the chain. This is sad since in the original paper FlyClient is advertised as a super-light client with sublinear chain weight proof that contains only $O(\log_2{L})$ headers together with inclusion proofs. However, we observe that the client which directly verifies every difficulty transition has neither polylogarithmic overhead nor sublinear overhead. It's nothing but a casual client with linear overhead. Nevertheless, B{\"u}nz \textit{et al.} claim that the specific variable-difficulty MMR described in the subsection 6.1 can resolve all issues at once. So let's move from the question (a) to question (b).   
 %"Bitcoin’s transition rule can be expressed as a parameterization of the variable difficulty model."
%If difficulty adjustment occurs every $N$ blocks then some headers at the ends of the interval should be downloaded together with inclusion proofs. 
%For the sufficiently large $N$ this approach could bring an improvement. 
%In aforementioned paper Ethereum is considered as a candidate for such improvement. However, in Ethereum the difficulty adjustment occurs every block. It means the original design has a serious compatibility issues with some existing solutions.

In the subsection 6.1 B{\"u}nz \textit{et al.} list a series of checks which are "sufficient for the simplified scenario". However, one can notice that this list covers only a part of calculations made in the difficulty transition protocol. Instead of the full check it verifies whether there is a possible assignment to the difficulty transitions yielding these parameters. Verification covers only the part of MMR that was included in the proof. Moreover, verification of the particular block difficulty often requires timestamp data from the previous epoch. So only a part of difficulty transition are actually verified. In order to finish the proof the authors of the original design address this problem with the following claim:

\begin{quote}
"We formally prove this by saying that an adversary that uses invalid difficulty transitions cannot increase its success probability."
\end{quote}

It's followed by the Lemma 4. It's time to move from question (b) to the question (c).

We have already discovered that a reliable verification requires a transmission of the significant amount of data from the original chain. Another argument against verification of difficulty transition rules is the fact that it doesn't prevent the malicious actor from manipulating timestamps. Attacker who secretly generates a concurrent chain can set timestamps as he wants. Therefore he can manipulate block difficulties in his favor while the resulting chain will respect the difficulty transition rules. Interestingly, the similar statement was proved by B{\"u}nz \textit{et al.} in Lemma 4.

\subsubsection{Investigation of the motivation}
Let's investigate why aforementioned verification was established. The author of the FlyClient claims that this verification is required to prevent the difficulty raising attack introduced by Banack 
%\textit{et al.}
\cite{bahack2013theoretical}. In particular, in the introduction on the page 4 it's written that

\begin{quote}
    "Without these checks an adversary could create valid proofs with non-negligible probability by creating few but high difficulty blocks."
\end{quote}

This statement is followed by the "SPV assumption":

\begin{assumption}(SPV assumption). 
The chain with the most PoW solutions follows the rules of the network and will eventually be accepted by the majority of miners.
\end{assumption}

Also on the page 7 one can read: 

\begin{quote}
    "The main intuition for these attacks is that an adversary can mine fewer but higher difficulty weight blocks such that, because of the increased variance, they can get lucky and exceed the more powerful honest miners."
\end{quote}

In fact, the model chosen by the author of FlyClient improperly represents commonly accepted views on the Proof-of-Work consensus and SPV clients. 

In the model chosen by B{\"u}nz \textit{et al.} the "correct" chain is one that is longer. The role of the longest chain rule in the original FlyClient design is set in Algorithm 1 in the subsection 4.1. However, the correct chain should be the one in which more work was executed. Aforementioned work is the aggregate difficulty of the blocks in the chain. The author of the FlyClient address the problem by introducing "SPV assumption". Actually, this assumption is wrong. It's a well known confusion about the Proof of Work consensus.\cite{lengthVSweight} %Although this confusion is well known in the community of academics and 
This incorrect implementation of the Proof of Work consensus opens the door to multiple attacks based on timestamp manipulation.\cite{lengthVSweight} It appears that it's very easy to construct malicious chain that is longer than the heaviest chain maintained by the honest majority. Moreover, difficulty transitions within it will be correct. The well known example is Bitcoin's testnet3 chain. It suffered from such attacks and became much longer than Bitcoin's chain in the main network.

\subsubsection{Investigation of Banack's draft}
In the FlyClient original paper aforementioned attacks are ignored. On the other hand, the difficulty raising attack is mentioned. However, this attack has a wrong evaluation. %It was misused, and the argument written in the paper is misleading.

\begin{enumerate}[1)]
%    \item 
%\begin{itemize}
\item
First, the paper by Banack is a draft written in 2013. There are some incomplete arguments in it.
%\item
%Second, the difficulty raising attack proposed by Banack doesn't match the one discussed by by B{\"u}nz \textit{et al.}
\item
%Third, 
The paper written by Banack has nothing to do with the security of light clients and SPV protocol. It describes the interaction between full nodes maintained by miners. Some important details of consensus rules are missed.
%\item
%Fourth, honest miners in Banack's model don't follow commonly accepted PoW consensus rules. It's likely that the consensus model based on the longest chain rule was used. The problem is that aforementioned rules were never listed within the paper. Therefore, it's hard to evaluate the relevance of the results presented in the paper.
\item
Finally, similarly to the St. Petersburg Paradox\cite{martin2011st} in Probability Theory, effectiveness of Banack's attack is limited by real life restrictions. In particular, the duration of any attack is always limited. Common PoW consensus rules reject forks which occurred deep in the chain history. Probability of success and expected profit of the attack is negligible compared to the significant resources required for its launch. This attack can harm. However, casual 51\% attacks appear to be more affordable and profitable for malicious actors.  
%\end{itemize}
\end{enumerate}

\subsubsection{Fixing the problem}
One may notice that this attack is similar to martingale strategies in casino games. The simplest way to prevent such attacks is to set an upper bound on the block difficulty. The natural bound is the difficulty of the blocks on BC taken with some coefficient. So we can set the rule that the difficulty of blocks on every SC should be less than maximal difficulty of blocks on BC within some interval divided by 20. Super Light Client should reject shard chains with extra difficult blocks. Full nodes should follow this rule too.

\textbf{Super-light client Rule 1.} Nodes that run shard client in sharded blockchain network should reject SC block which violate the restriction on the difficulty:
\begin{equation}
D(\text{SC block}) \le C \cdot \max_S(\text{BC block})
\end{equation}
Where $C$ is some coefficient and $S$ is some subset of blocks on the main chain.

%One might argue that this rule doesn't work in the casual case when there is no shards and no BC to get a BC difficulty for the reference. In this case we can set another rule. It should prevent the super-light client from accepting malicious chain with the high variance. 

%\textbf{Super-light client Rule 2.}

\subsection{Problem with the optimal sampling}
Interestingly, the problems in the FlyClient original design\cite{bunz2019flyclient} are not limited to the determination of the heaviest chain. There are numerous issues in subsections 5.4 and 6.1. Some problems tricky and require a thorough discussion like the one above. The attempt to define and determine the best probability distribution for sampling is a complete fail. Some errors are rather plain. Description, discussion and resolving these issues goes beyond the scope of this subsection in the Appendices. It's the subject of a subsequent paper.

According to the Google Scholar search engine the aforementioned paper by B{\"u}nz \textit{et al.} was already cited in at least 7 other papers. Interestingly, there is no mention about any issues within those papers. Moreover, the aforementioned papers assert that FlyClient is implemented for Ethereum. However, in the original paper it is clearly stated that FlyClient has a serious compatibility issues with Ethereum blockchain. Therefore FlyClient was proposed for "Ethereum-like blockchain" which has significant differences with real Ethereum blockchain. So these assertions about the FlyClient are inaccurate and misleading.

%There are some problems with sampling a subset of blocks that constitute the MMR proof. As it is discussed in the original paper, the optimal strategy of the malicious actor is to fork the honest chain at some block and construct his own fork. This malicious fork consists of some number of invalid blocks and some number of blocks correctly mined by the malicious actor. Due to the properties of the MMR malicious fork doesn't contain

%as it's stated in the subsection 6.1, to enable handling difficulty-based sampling, there is a need to make an adjustments. "We need a data-structure which efficiently and securely enables the verifier to sample blocks based on their relative difficulty positions, rather than their absolute positions as in the standard MMR."

%squize by 2, squize by 2
%log log log

%2) Он пишет "возьмем функцию плотности вероятности"
%Если говорить математически, Генерация блоков это вообще говоря стохастический процесс. Когда цепочка сгенерирована "вес блока" это дискретная вероятностная мера

%In the paper \cite{} Karakostas \textit{et al.} in the Table 1 on the page 20 claims that FlyClient is already implemented in the Ethereum. 

\section{Reward scheme details}
\subsection{Complexity of the mining in the blockchain} \label{ComplexityOfM}

The goal of this subsection is to set rewards for mining in correspondence with the computational time spent for mining so that the miner who computed $X$ megahashes will receive the proportional reward measured in coins. At list we want make the fraction between reward and computational time to be as close to constant as possible. Computational time here is not the time spent for mining by certain mining farm but the number of hashes which were computed.

\begin{comment}
Let's consider Bitcoin blockchain\cite{Satoshi} as a base for our construction. First, let's find the complexity of generating a certain block in a classical Bitcoin network. Bitcoin blockchain uses Proof of Work (PoW) paradigm so miners mine blocks with specific properties in order to get a reward. The block structure \cite{BlockStr} as December 2016 consist of following parts:

\begin{table}[ht]
	\centering
	\captionsetup{width=.85\textwidth,font=footnotesize, justification=centering}
	\caption{BTC Block structure \label{BTCstruct}}
	\renewcommand{\arraystretch}{1.5}%
	%\resizebox{\textwidth}{!}{
    \begin{tabular}[center]{|c|c|}\hline
	Content in block & Size in bytes\\ \hline
	Magic Number & 4\\ \hline
	Block size & 4\\ \hline
	Version & 4\\ \hline
	Previous block hash & 32\\ \hline
	Mercle root & 32\\ \hline
	Time stamp & 4\\ \hline
	Difficulty Target & 4\\ \hline
	Nonce & 4\\ \hline
	Transaction counter & 1-9\\ \hline
	Transaction List & Upto 1MB\\ \hline
    \end{tabular}\\
    %}
\end{table}
\end{comment}

Let's consider the block content described the \cref{BlockCont}. Block headers on BC and SC contain a specific field often called Bits. It encodes a positive integer \textit{target difficulty} or simply \textit{target}. Even though it is a $256$ bit number, $32$ bits and some formula are used to compress and store it in the block.
%Target difficulty number \cite{BlockMine} is of particular importance for us. 
During the mining process the miner composes the block and calculates its $256$ bit mining hash which is discussed in the \cref{HashFunctions}. If the hash considered as binary number is less than target difficulty number then the mining was successful.

%The target adjusts every $2016$ blocks. Approximately that is once a two weeks. 

Let's denote the target as $T$. Then the probability $p$ to get low hash after single hash computation is:
\begin{equation}
    p=\frac{\text{number of low hashes}}{\text{total number of hashes}}=\frac{T}{2^{256}}
\end{equation}

We can consider single block mining as a classical Bernoulli process in which we repeat the block generating procedure until we find the block with a good hash. Here the success event is "the low hash found" and failure event is "the block hash is to too high". We don't know exactly when the success event will occur, however, mathematical expectation of the number of blocks generated until the success is equal to $\frac{1}{p}=\frac{2^{256}}{T}$. So target $T$ is a good measure of the current complexity of mining in the blockchain.

Here we should place an important disclaimer. In this section we consider an ideal situation when all miners in the network are honest and follow the prescribed protocol and timely share information with other nodes. We only calculated the probability to mine the block but not the probability to get the reward for the mining. In reality the blockchain may have collisions when two miners simultaneously mine blocks. In this case only one of the blocks will be placed into the chain according to the consensus protocol accepted in the network. Therefore, only one of the miners will get a reward. Moreover, some of the nodes may perform so called "selfish mining" and increase their reward by some factor. Many author describes a scenario in which a selfish miner or some coalition of them avoid sharing information with honest nodes. In Jaxnet we use the fix proposed in the paper \cite{Eyal_2018}.
%This actions maybe unprofitable and risky unless the coalition of selfish nodes has more then $\frac{1}{3}$ of network computation power. However, in the later case selfish miners get the significant advantage.   
\subsection{Setting rewards for the mining}\label{SetReward}

Suppose we want to set a reward in coins for each mined low hash in such a way that on average the reward is proportional to the effort made by the mining network. Obviously, it's not efficient to track the performance of every miner, every mining farm and every Application-Specific Integrated Circuit (ASIC). However, according to the \cref{ComplexityOfM} we have an estimate for it. That is
\begin{equation}
    \frac{2^{256}}{T}
\end{equation}
On average that is a total number of blocks mined by the whole network before somebody will find the good hash.

In the classic Bitcoin blockchain, the reward remains fixed for a long time interval. Then, the block mining reward in this network halves every $210000$ blocks. At this moment we ignore the fees collected by miners in choosing which transaction should be included to the next block. Our goal is to set mining reward proportional to the mining complexity $\frac{T}{2^{256}}$ of the block. Suppose the reward function $R$ is not fixed and depends in some way on the quality of the hash of the block.
\begin{equation}
    R(\text{ block }) \neq \text{const}
\end{equation}
In terms of statistics our goal is to have the mathematical expectation of the reward for the mining pool, which is proportional to the complexity of mining. It is estimated as the mathematical expectation of the number of hashes to be mined before the success.
\begin{equation}
    E(R(\text{ block }))=k\cdot E(\text{ numb of blocks }) = k \cdot \frac{2^{256}}{T}
\end{equation}

where $k$ is some constant coefficient. The probability space here is the set of all possible mining events which end up with generating some low hash block. This good block is denoted as "block". Although, this block is not unique and it is even possible to have two good blocks to be generated simultaneously. We will discuss this situation later in this paper. We assume that the hash function is perfect and all good hashes and all good blocks are equally likely to be generated.

The shortest way to rich this goal is to set the reward function equal to $k\cdot \frac{2^{256}}{T}$. In this case
\begin{equation} \label{eq:VanillaRew}
    E(R\text{( block )}) = R(\text{ hash }) = k\cdot\frac{2^{256}}{T}
\end{equation}
and everything is fine. 

Let us discuss the advantages of this approach for awarding rewards. The main effect of reward distribution in this way is that we set a coin issuance to be dependent on hash rate of the network which mines the chain. This effect is more valuable when we have a network with multiple shards. Each shard of the network perform mining and coin issuance. If both shards issue coins with the rate proportional to their hash rate we could set a correspondence between the value of coins in the first and in the second chain so that we can exchange in one to one rate. The value of the coin will be in direct correspondence to the amount of computational resources involved in mining of blocks in particular time interval. In particular every miner in the network will know that no matter what chain he will choose for mining he will get the same expected reward of issued coins. Really, if the miner $M$ has a hash rate $x$ of the hash rate of the network then the probability that he will get the reward $R$ is $x$. Thus his expected reward is $xR$. On the other hand the expected number of hashes that he will calculate during the mining round is $x$ part of the expected number of hashes given that will be calculated in the whole network during the mining round. In our model this expected number of hashes calculated during the round is proportional to the reward. Thus, the expected reward of each miner is proportional to his effort. However, there maybe different variance of expected reward on each chain. In our model the chain with higher hash rate will get lower parameter target and therefore higher complexity of mining the block. If a miner with low hash rate mines on that chain he will get rewarded less often than on a chain with lower complexity. Therefore, his income from mining on chains with high target will be more stable.

Another part of miner's reward is transaction fees. Any miner will be interested to direct computational power into mining in the chain which has a higher ratio
\begin{equation}
    \frac{\text{aggregate transaction fees}}{\text{aggregate hash rate of miners on the chain}}
\end{equation}
Therefore, the higher number of aggregate transaction fees will stimulate miners to mine next block on this chain. On the other hand, if there are less transaction requests, mining and block creation will go slower. Similar to Bitcoin blockchain after some number of of blocks the difficulty parameter "target" will be adjusted in order to stabilize the block creation rate in the network.

The separate case is when there is a collision in the network and the chain temporary splits into two chains. Then the consensus mechanism comes into play and decides which chain will become a main chain. Those who were mining this chain will get the reward by collecting transaction fees and some number of issued coins. The question is whether those miners who were mining on the second chain or even third chain get any reward. In the Bitcoin network they don't get anything and blocks in it become "orphan blocks" not used in the network. In the Ethereum network, there is another approach. Blocks outside of the main chain could become so-called uncle blocks. Uncle blocks get some part of the reward. The second approach has some advantages with some minor disadvantages. 
%We are going to discuss them in more detail in the next chapter.

One important observation here is that there is not much sense in giving high part of reward to the miners who mine blocks outside of the main chain even though later they will be incorporated into the chain. It's crucial to maintain the main chain and thus have the consensus on what transaction were processed in the network. Even in more complex protocols proposed or used in Ethereum there is an analogue of the main chain which has just more complicated building rules. Thus its crucial to stimulate miners to mine on the main chain, and in the case of collision, always direct their resources to more attractive chains. In the case of Bitcoin blockchain the more attractive chain is one which is longer. Giving more reward to orphan or uncle blocks may stimulate selfish mining and rise security concerns. On the other hand if we want to make the blockchain network with large number of transactions and low transaction processing delays we need to have high block creation rate. When the block creation rate increases collisions in the network occur more often. Each such collision involves a certain amount of mining by some part of the miners outside the main chain. Moreover, even in the case when there are no collisions, mining outside the main chain still occurs shortly after the new block creation due to information transition delays. The Internet network may be fast; nevertheless the speed of light is considered to be a natural limit to conventional ways of transmitting information in time-space. If a miner on average has a half second delay before he learns that the new block was created in the network, then on average he will spend half a second on mining outside of the chain each round. For example, a Bitcoin network with 10 minute rounds will result in 3 seconds of mining time during each hour. If new blocks are created every 30 seconds then wasteful mining will take 1 minute per hour. Thus the aggregate time which miners spent on mining outside the main chain significantly increases. %Therefore it's natural to set more fair reward distribution algorithm by assigning part of issued tokens to uncle blocks. 

The essential question is whether the miners get rewarded proportionally to their effort when a collision has occurred on the chain in our model. The answer is as follows. Whenever the miner mines on the main chain his expected reward is proportional to his effort. We assume that the mining community in the network follows the protocol, and do not try to perform $51\%$ attack or any other cheating, so that whenever a new block is created earlier than its counterparts, and it was immediately transmitted to the network, then with overwhelming probability it will be adopted by the network, and the miner will get his reward. Of course, it's possible that the miner has high ping, unreliable internet connection, low karma, bad luck, or whatever else, and his block won't get into the chain. Nevertheless, we assume that such situations won't be common, and it's in each miner's interest to reduce off-chain mining impact on their income. Our model suggests some part of reward for uncle blocks. Nevertheless, we expect that mining would be almost always performed on the main chain. Therefore the paradigm "expected reward equals effort" will hold up to small amendment described above. It worth saying that such collisions are an essential part of all blockchain designs and such assumptions are an essential part for all of them. One of the goals of our proposal is to adjust the parameters of our model in the way that reduces mining time loss due to occasions described above.

\section{Economic details}
\subsection{Economic model in Jaxnet}\label{EconomicModel}
\subsubsection{Assumptions about money and monetary rules} \label{Economic assumptions}
Let us make few assumptions about money and monetary rules before delving into the economic setup. We follow a micro-founded understanding of money, where its fundamental value converges towards its transaction purpose. \cite{tirole1985asset} Our coin supply is endogenous, i.e. it depends on economic variables and not on an external authority. However, coin supply does not abide by the same rules as in standard monetary economics. For one thing, as liquidity increases into the economy, the price levels decreases. More precisely, the coin value denominated in other currencies decreases. 

Misleadingly, some economists have argued that an equilibrium exists where the value of a privately issued outside money such as Bitcoin is zero \cite{garratt2018bitcoin}. But this relies on two strong assumptions:
\begin{enumerate}[i)]
\item
marginal costs of minting are decreasing and,
\item
miners never adjust their supply to market conditions.
\end{enumerate}
If i) can be true with paper money printing, this is not necessarily the case in the cryptocurrency world. In our setting, mining one extra coin requires more electricity (variable cost) and more computing power (irreversible investment as in \cite{dixit1994investment,prat2018equilibrium}), provided the mining pool is already at full capacity. Mining cryptocurrencies is costly, unlike printing banknotes. Furthermore, we argue that miners are price takers who compete on quantities, exactly like in a Cournot setup \footnote{In this setup, assume that there are several mining pools competing with each other. They cannot compete on the prices of the output (i.e. JaxCoins), since this one is given by market forces. Instead, they adjust their quantities (here the hash rate) according to the quantities of the other mining pools in the market.}, where they adjust the quantity of hashpower according to others’ choices. Therefore, global production depends only on the sum of marginal costs \cite{varian2014intermediate}. This is confirmed by one econometric analysis, where the price of Bitcoin is closely correlated to its marginal cost of production \cite{hayes2019bitcoin}. In that case, money supply is a function of its underlying cost structure (i.e. mining intensity) on the supply side and the demand for transnational purposes on the other \cite{biais2018equilibrium}. 

Overall, money supply adjustments in our framework follows a sequence such as:
\insertimage{DecisionTimeline.png}{0.95}
Where:
\begin{itemize}
\item
$p$ is the price of outputs
\item
$w$ is the price of inputs
\item
$\pi$ is the profit
\item
$M$ is the monetary mass, i.e. all coins in circulation
\item
$T$ is the number of transactions over the network
\item
$V$ is the velocity of JaxCoin
\item
$E[v]$ is the expected value of JaxCoins for the next period
\end{itemize}

Here, we can see that the demand and supply adjust sequentially. Output (i.e. new coins minted) are always an ex post response to higher or lower demand for JaxCoins. Miners adjust their behaviors to market conditions. Let us suppose there is a drop in demand for transactions on the network. After this drop, miners will calculate their expected profit for the next period and allocate the proper amount of computing power that maximize their profits (i.e. market price of the total amount of coins they can mine minus their costs). In these conditions, miners always adjust to demand shocks with one period time lag. 

Therefore, this figure shows that there is a feedback loop that can amplify the impact of demand shocks in the short run. In the particular case where the users’ expected value drops, they will ditch the coin and drive its value even further down. This is reinforced by the fact that once created, newly minted coins will be circulating forever and be part of the monetary mass, unless otherwise lost. Miners will adjust and lower their hash rate to maintain their marginal profits equal to their marginal cost. The converse being true during a period where demand increases.

Once again, miners are profit driven. They make money in allocating their computing power. Therefore, the expected value of the coin is directly linked to its marginal cost, which, we assume, would limit long term volatility. 
\subsubsection{Short term risks for JaxCoin}\label{ShortTermRisks}

At early stages, prices will not be denominated into Jaxcoins as it needs first to be considered as a medium of exchange. Inflation concerns are set at a more mature stage of Jaxcoin history. Furthermore, the implementation of smart contract functions into the Jax.Network will mix things up, as Jaxcoin will not be only a medium of exchange but also a utility token . 

That being said, the primary concern for JaxCoin is a currency crash. Assuming JaxCoin will follow a logit rate of adoption, prices will not be denominated in JaxCoin at early stages. At this stage, the platform still needs to attract a sustainable network effect in a two-sided setup. Thus, our concern is that the currency would not be stable because of oversupply and will depreciate against other fiat or cryptocurrencies. 

In modern times, Reinhart \& Rogoff “define as currency crash an annual depreciation in excess of 15 percent. Mirroring our treatment of inflation episodes, we are concerned here not only with the dating of the initial crash but with the full period in which annual depreciations exceed the threshold.” \cite{reinhart2009time}

A GPU/ASIC-based supply would increase the money supply between 18 to 80\% year on year, according to our worst-case scenario. Indeed, the efficiency gains of mining equipment are forecast to increase by this rate. This will affect primarily the exchange rate, as the coin will first be denominated in other currencies or cryptocurrencies. Only in a second time-period, after the adoption diffusion has reached a certain level, that some inflation can arise. In both cases, we propose some rules to overcome these issues. 
\subsection{Inflation of the coin price} \label{PrInfl}

In this section we are going to discuss the potential decrease of the coin's purchasing power in our system. In short, we expect that the amount of goods that one user can  buy with $N$ coins could decline after some interval of time. However, we argue that this level of inflation will remain low and do not last in the long run.

There are a few reasons which make this thing happen. First, whenever a new block is created some amount of coins are issued in order to reward miners. These newly issued coins are added to the total amount of coins already in the system. Therefore, the total supply of coins grows up. If the demand on the coins grows at the same rate then purchasing power of the coin remains the same \footnote{It is assumed here that the velocity of the coin is constant. However, velocity has an impact on price levels. Indeed, it depends on the behavior of token holders. For instance, if they are willing to hold the coin for a long time period, velocity will be low. This means that more coins are needed to fill the transaction needs of the network. Unfortunately, velocity is very complicated to measure, and we do not wish to provide inaccurate forecasts on this account}. However, if the issuance of coins exceeds the demand of coins its price should go down. Second, we know that efficiency of mining equipment improves every year. For example, Cambridge Bitcoin Electricity Consumption Index (CBECI) uses the following data\cite{CBECI} on ASIC efficiency in order to estimate aggregate power consumption of the Bitcoin network.
%\insertimage{CambridgeISICindex.png}{0.9}
Also the aggregate hash rate of miners grows, mostly due to new entrants. This implies that the higher price for the coin the more profitable its mining and more miners will target their equipment to mining JaxCoin. Since the rate at which our coin is issued is proportional to the effort of miners, soon we will get a bunch of new coins in the system. These new coins will push its exchange rate downwards. Therefore, we do not expect the price of the token to be significantly higher than its mining cost in the long run. Furthermore, low price will reduce the mining activity in the network and slow down the coin issuance. In this scenario the main source of profit is stemming from mining fees. So, cost efficiency of mining plays a crucial role in adjusting the supply upward or downward. 

In order to control monetary issuance and limit supply shocks, we need to get an estimate of the efficiency gains we have to study how rapidly grows the efficiency of mining hardware. In this context one might remember the so-called Koomey's law.\cite{koomey2010implications} It states that from 1946 to 2009 power efficiency of microprocessors was doubling every $1.57$ years. That is $55.5\%$ of efficiency gains per year. However, in a subsequent work\cite{koomey2015moores}, this data was reexamined. It states that in the new century efficiency gains might be slowing down. According to the new estimate the doubling occurs every $2.7$ years. So, in 2015 there were $29.3\%$ of efficiency gains on average year on year. Although this data is somewhat outdated, there is sufficient thorough research on this topic and the current state of things is not clear. Moreover, technological progress in Application-Specific Integrated Circuits (ASIC) manufacturing tends to be more rapid than in General Processing Unit (GPU) manufacturing. Nevertheless, the recent research by Yo-Der Song and Tomaso Aste \cite{song2020cost} 
%sheds the light on this issue. 
that study efficiency improvements in the mining hardware from 2010 to 2020.
However, the methodology in this research differs from our approach. 
%The dataset in it includes the history of oil prices. Moreover, 
GPU mining and ASIC mining are not distinguished in the calculation. Therefore, the estimate for the technological progress given in this study doesn't accurately address the needs of our model.

Let us study the efficiency gains of ASICs. Table \ref{ASICchart} shows that the list of efficiency leaders in past $5$ years was the following:
\begin{table}[ht]
\centering
\captionsetup{width=\columnwidth,font=footnotesize,
  justification=centering}
  \caption{ASICs efficiency leaders timeline}
  \label{ASICchart}
{
\def\arraystretch{1.5}%
\resizebox{0.95\columnwidth}{!}{
\begin{tabularx}{\columnwidth}{@{\extracolsep{\fill}}Xll}
\toprule
%\hline
Name of the ASIC & Release date & Efficiency (J/Gh)\\ %\hline
\midrule
Bitmain Antminer S3 & 07.2014 & 0.77 \\ %\hline
Bitmain Antminer S3 & 12.2014 & 0.51 \\ %\hline
Bitmain Antminer S7 & 09.2015 & 0.27 \\ %\hline
Bitmain Antminer S9 (11.5Th) & 06.2016 & 0.1 \\ %\hline
Ebang Ebit E10 & 02.2018 & 0.09 \\ %\hline	
ASICminer 8 Nano Pro & 05.2018 & 0.05 \\ %\hline
Bitmain Antminer S17 Pro (53Th)	& 04.2019 & 0.04 \\ %\hline
\bottomrule
\end{tabularx}}
}
\end{table}
For the $62$ month-period between July 2014 and August 2019 we can determine ASIC with the best efficiency per Joule. We assume that efficiency of ASICs could be simulated as Geomtric Brownian motion. The common approach in this setting is to consider logarithms of efficiency of ASICS and approximate them in least square metric with a linear function of time.

Unsophisticated computation in Octave gives us the estimated coin growth rate of $80,35$ percents year on year or $5$ percents per month.
\insertimage{Octave1.png}{0.8}

%\insertimage{OctaveRES1.png}{0.6}
%\insertimage{ASICeff.png}{0.9}
Other studies of ASIC's efficiency could be found here \cite{BTCVice,BTCefftoDouble}.

Of course, it is doubtful that this level of efficiency could be achieved by any mining farm. First, buying and maintaining mining equipment costs huge amount of money. Second, besides mining hardware there is network hardware. It costs money and consumes electricity. Its efficiency vary from one data-center to another. Third, some mining farms may have access to cheaper electricity. For instance, they purportedly utilize the heat from mining rigs. 

Let us now turn to GPU mining.
\begin{table}[ht]
\centering
\captionsetup{width=\columnwidth,font=footnotesize,
  justification=centering}
  \caption{GPU hashes per Joel efficiency table}
  \label{GPUeffchart}
{
\def\arraystretch{1.5}%
\resizebox{0.9\columnwidth}{!}{
\begin{tabularx}{\columnwidth}{@{\extracolsep{\fill}}>{\raggedright\arraybackslash}Xlllll@{}} 
\toprule 
%\hline
\makecell{Name of \\ the GPU} & 
\makecell{hash \\ algorithm} & 
\makecell{Release\\ date} & 
\makecell{hashrate \\ Mhashes/s} & 
\makecell{Power \\ consumption W} & 
\makecell{Efficiency\\ (Mh/J)}\\ \hline
%HD 4870 & poclbm & 06.2008 & 0.872 \\ \hline
%HD 4770 & - & 04.2009 & 0.904 \\ \hline
Rx 380 & Ethash & 06.2015 & 19 & 140 & 0.219 \\ %\hline
Rx 390 & Ethash & 07.2015 & 16.5 & 220 & 0.075 \\ %hline
Rx 480 & Ethash & 06.2016 & 29.5 & 135  & 0.136 \\ %\hline
Rx 580 & Ethash & 04.2017 & 30.2 & 135 & 0.224 \\ %\hline
Vega 56 & Ethash & 08.2017 & 36.5 & 210 & 0.174 \\ %\hline
Radeon VII & Ethash & 02.2019 & 78 & 230 & 0.339 \\ %\hline
GTX1060 (6GB) & Ethash & 07.2016 & 22.5 & 90 & 0.25 \\ %\hline
1070 & Ethash & 06.2016 & 30 & 120 & 0.25 \\ %\hline
1660TI & Ethash & 02.2019 & 25.7 & 90 & 0.286 \\ %\hline
2060 & Ethash & 01.2019 & 27.6 & 130 & 0.212 \\ %\hline
2070 & Ethash & 10.2018 & 36.9 & 150 & 0.243 \\ %\hline
2080TI & Ethash & 09.2018 & 52.5 & 220 & 0.239 \\ %\hline
\bottomrule
\end{tabularx}}
}
\end{table}

Again we take the logarithm of the efficiency and apply a linear approximation in least square metric. Unsophisticated calculations in Octave give us inflation of $18$ percents per year.
\insertimage{Octave2.png}{0.8}
%\insertimage{OctaveRES2.png}{0.6}

In either case, we just provided a rough approximation as we lack data points to be more precise. This is due to the fact this market is quite new and we do not have much hindsight on the potential efficiency gains in the long run, especially with ASIC circuits. 

One might question why other mining expenses are not taken into account. As we know from the \cref{table:BTCcost}, expenses on mining hardware and electricity are much higher than other expenses. The price of the certain mining appliance is not fixed. The highest price on the new model of ASIC is within first few weeks after its release on the market. Then its price dramatically deteriorate until power efficiency of the hardware becomes insufficient to compete with new models and keep mining profitable. We can conclude that the price of the ASIC completely depends on the profit they could bring during there lifespan. Interestingly, the most important property of these equipment is power efficiency.

One might question why we don’t use the data about hash rate dynamics and USD-cryptocurrency exchange rates. We posit that cost of hash computation can be derived from this data. The problem is that the price of the cryptocurrency is highly speculative and volatile. The price can plummet within a short period of time. However, these ups and downs do not imply that mining hardware becomes worth in hash computations.

\end{document}